\documentclass[12pt]{article}

\usepackage{amstext, amssymb, amsthm, amsmath,bm,graphicx,rotating,array,natbib,stmaryrd,booktabs,color}
\usepackage{algorithmic,algorithm}
\usepackage{multirow}
\usepackage{enumerate}
\usepackage{graphicx}
\usepackage{caption}
\usepackage{subcaption}
\usepackage{hyperref}
\usepackage[total={6.2in,8.75in}]{geometry}
\newtheorem{proposition}{{\bf Proposition}}

\newtheorem{definition}{{\bf Definition}}
\newtheorem{theorem}{{\bf Theorem}}

\newcommand{\var}{\mathrm{var}}
\newcommand{\avar}{\mathrm{avar}}
\newcommand{\cov}{\mathrm{cov}}
\newcommand{\vect}{\mathrm{vec}}
\newcommand{\vech}{\mathrm{vech}}

\newcommand{\indep}{\;\, \rule[0em]{.03em}{.6em} \hspace{-.25em}
\rule[0em]{.65em}{.03em} \hspace{-.25em}
\rule[0em]{.03em}{.6em}\;\,}
\newcommand{\real}[1]{\mathrm{I \! R} \mathit{^{#1}}}
\newcommand{\trans}{^{\mbox{\tiny {\sf T}}}}
\newcommand{\spn}{\mbox{span}}
\newcommand{\spc}{{\mathcal S}}
\newcommand{\cspc}{{\mathcal S}^{\tiny \perp}}

\newcommand{\Abf}{{\bm A}}
\newcommand{\Bbf}{{\bm B}}
\newcommand{\Cbf}{{\bm C}}
\newcommand{\Dbf}{{\bm D}}

\newcommand{\Gbf}{{\bm G}}
\newcommand{\Hbf}{{\bm H}}
\newcommand{\Ibf}{{\bm I}}
\newcommand{\Jbf}{{\bm J}}
\newcommand{\Kbf}{{\bm K}}

\newcommand{\Mbf}{{\bm M}}
\newcommand{\Nbf}{{\bm N}}
\newcommand{\Obf}{{\bm O}}
\newcommand{\Pbf}{{\bm P}}
\newcommand{\Qbf}{{\bm Q}}

\newcommand{\Ubf}{{\bm U}}
\newcommand{\Vbf}{{\bm V}}

\newcommand{\Xbf}{{\bm X}}
\newcommand{\Ybf}{{\bm Y}}
\newcommand{\Zbf}{{\bm Z}}

\newcommand{\cbf}{{\bm c}}
\newcommand{\ebf}{{\bm e}}

\newcommand{\gbf}{{\bm g}}
\newcommand{\hbf}{{\bm h}}

\newcommand{\sbf}{{\bm s}}

\newcommand{\wbf}{{\bm w}}

\newcommand{\zerobf}{{\mathbf 0}}

\newcommand{\greekbold}[1]{\mbox{\boldmath $#1$}}

\newcommand{\betabf}{\greekbold{\beta}}
\newcommand{\deltabf}{\greekbold{\delta}}

\newcommand{\phibf}{\greekbold{\phi}}

\newcommand{\varepsilonbf}{\greekbold{\varepsilon}}

\newcommand{\thetabf}{\greekbold{\theta}}

\newcommand{\Deltabf}{\greekbold{\Delta}}
\newcommand{\Gammabf}{\greekbold{\Gamma}}
\newcommand{\Thetabf}{\greekbold{\Theta}}
\newcommand{\Sigmabf}{\greekbold{\Sigma}}

\newcommand{\Omegabf}{\greekbold{\Omega}}

\newcommand{\xibf}{\greekbold{\xi}}

\newcommand{\Ptensor}{{\mathbb P}}
\newcommand{\Qtensor}{{\mathbb Q}}

\newcommand{\bhatols}{\widehat \betabf_{\textrm{{\tiny OLS}}}}
\newcommand{\bhatenv}{\widehat \betabf_{\textrm{{\tiny ENV}}}}
\newcommand{\Bhatols}{\widehat \Bbf_{\textrm{{\tiny OLS}}}}
\newcommand{\Bhatenv}{\widehat \Bbf_{\textrm{{\tiny ENV}}}}
\newcommand{\Bhatenvit}{\widehat \Bbf^{it}_{\textrm{{\tiny ENV}}}}
\newcommand{\Bhatenvos}{\widehat \Bbf^{os}_{\textrm{{\tiny ENV}}}}

\newcommand{\Sigx}{\Sigmabf_{{\tiny \Xbf}}}

\newcommand{\env}{{\mathcal E}}
\newcommand{\envsk}[1]{\env_{\bm\Sigma_{#1}}\left( \Bbf_{(#1)} \right)}
\newcommand{\PGammak}[1]{\Pbf_{\mathbf{\Gamma}_{#1}}}
\newcommand{\envs}{\env_{\bm\Sigma}(\Bbf)}
\newcommand{\envt}{{\mathcal T}_{\bm\Sigma}(\Bbf)}

\title{\Large{\textbf{Parsimonious Tensor Response Regression}}}
\author{
\bigskip
Lexin Li and Xin Zhang\\
\normalsize{University of California, Berkeley; and Florida State University}
}
\date{}

\begin{document}
\maketitle

\begin{footnotetext}[1]
{Lexin Li is Associate Professor (Email: lexinli@berkeley.edu), Division of Biostatistics, University of California, Berkeley, Berkeley, CA 94720-3370.
Xin Zhang is Assistant Professor (Email: henry@stat.fsu.edu), Department of Statistics, Florida State University, Tallahassee, FL
32306-4330.}
\end{footnotetext}

\baselineskip=21pt

\begin{abstract}
Aiming at abundant scientific and engineering data with not only high dimensionality but also complex structure, we study the regression problem with a multidimensional array (tensor) response and a vector predictor. Applications include, among others, comparing tensor images across groups after adjusting for additional covariates, which is of central interest in neuroimaging analysis. We propose parsimonious tensor response regression adopting a generalized sparsity principle. It models all voxels of the tensor response jointly, while accounting for the inherent structural information among the voxels. It effectively reduces the number of free parameters, leading to feasible computation and improved interpretation. We achieve model estimation through a nascent technique called the envelope method, which identifies the immaterial information and focuses the estimation based upon the material information in the tensor response. We demonstrate that the resulting estimator is asymptotically efficient, and it enjoys a competitive finite sample performance. We also illustrate the new method on two real neuroimaging studies.
\end{abstract}


\newpage
\section{Introduction}

For modern scientific data, an overarching feature that accompanies high or ultrahigh dimensionality is the complex structure of the data. For instance, in neuroimaging studies, electroencephalography (EEG) measures voltage value from electrodes placed at various scalp locations over a period of time, and the resulting data is a two-dimensional matrix where the readings are both spatially and temporally correlated. Similarly, anatomical magnetic resonance imaging (MRI) takes the form of a three-dimensional array, where voxels correspond to spatial locations of the brain and are spatially correlated. Multi-dimensional array type data also frequently arise in chemometrics, econometrics, psychometrics, and many other applications. In our study, we are primarily interested in comparing multidimensional array, also know as \emph{tensor}, under two or more conditions, after adjusting for other potentially confounding variables. Our motivating examples came from two neurological imaging studies, while the  proposed methodology equally applies to a variety of scientific and engineering applications as well. One example is to compare the EEG scans between the alcoholic subjects and the general population, and the second is to compare the MRI scans of brains between the subjects with attention deficit hyperactivity disorder (ADHD) and the healthy controls after adjusting for age and gender of the subjects. This tensor comparison problem can be more generally formulated as a regression with the image tensor as response and the group indicator and other covariates as predictors. In this article, we aim to study this problem and term it \emph{tensor response regression}.

While there is an enormous body of literature tackling regression with high- or ultrahigh-dimensional predictors, there have been relatively few works on regression with multivariate response, and even fewer works on regression with tensor response. We review three major lines of related research. The first concerns multivariate \emph{vector} response regression, and popular  solutions include partial least squares \citep{Helland1990, Helland1992, Chun2010}, canonical correlations \citep{ZhouHe2008}, reduced-rank regressions \citep{Izenman1975, Reinsel1998, Yuan2007}, sparse regressions with various penalties incorporating correlated response variables \citep{Simila2007, Turlach2005, Peng2010}, and sparse reduced-rank regressions \citep{Chen2012}. Most of existing solutions adopt a linear association between the multivariate response and predictors, and they universally deal with the case where the response variables are organized in the form of a vector. Our goal, however, is to model a tensor response, where the vector response can be viewed as a special case of a one-dimensional tensor. The second line of research directly models association between an image tensor and a vector of predictors in the context of brain imaging analysis. The dominating solution in the field regresses one response variable (voxel) at a time \citep{Friston2007}, and thus completely ignores underlying correlations among the voxels \citep{Li2011}. \citet{Li2011} and its follow-up works \citep{Skup2012,Li2013} proposed a multiscale adaptive approach to smooth imaging response and to estimate parameters by building iteratively increasing neighbors around each voxel and combining observations within the neighbors with weights. Our approach differs in that we aim to model all the voxels in an image tensor \emph{jointly} while incorporating the intrinsic spatial correlations among the voxels. Finally, there have been some recent developments regressing a scalar response on a tensor predictor \citep{ReissOgden2010,ZhouLiZhu2013, ZhouLi2014,Goldsmith2014,Wang2014}. By contrast, our proposal reverses the role by treating the image tensor as response and the vector of covariates as predictors. The two treatments yield different interpretations. The former, the tensor predictor regression, focuses on understanding the change of a clinical outcome as the tensor image varies, so may be used for disease diagnosis and prognosis given image patterns. The latter, the tensor response regression, aims to study the change of the image as the predictors such as the disease status and age vary, and thus offers a more direct solution if the scientific interest is to identify brain regions exhibiting different activity patterns across different groups of subjects. In addition, the technique proposed in this article is completely different from the techniques used in tensor predictor regression, and as we will later show in the simulations, tensor response regression exhibits a more competitive finite sample performance compared to tensor predictor regression when the sample size is small.

In this article, we propose a parsimonious tensor response regression model and develop a novel estimation approach. Specifically, we continue to impose a linear association between the tensor response and the predictors. Meanwhile, we adopt a form of sparsity principle by assuming that part of the tensor response does not depend on the predictors and does not affect the rest of the response either. Adopting this principle effectively reduces the number of free parameters, leads to a parsimonious model with improved interpretation, and yields a coefficient estimator that is asymptotically efficient. This principle can finds its natural counterpart in the sparsity principle in regression and variable selection with high-dimensional predictors, where only a subset of variables are believed to be relevant to the response. However, our proposal significantly differs from the popular sparse model estimation and selection in several ways. While the usual sparsity principle frequently adopted in variable selection assumes a subset of \emph{individual} variables are irrelevant, we assume that the \emph{linear combinations} are irrelevant to regression. Rather than using $L_1$ type penalty functions to induce sparsity, as is often done in variable selection, we employ a nascent technique called the \emph{envelope} method \citep{Cook2010} to estimate the unknown regression coefficient. Moreover, whereas most sparse modeling treats variable selection and parameter estimation separately, our envelope method essentially identifies and utilizes the material information in a joint estimation manner. We develop a fast estimation algorithm and study the asymptotic properties of the estimator. We demonstrate through both simulations and real data analyses that the new estimator improves dramatically over some alternative solutions.

The contributions of this article are multi-fold. First, it addresses a family of questions of substantial scientific interest but with relatively few solutions. Our proposal offers a systematic approach to jointly model all elements of a tensor response given a vector of predictors. A particular application is to compare tensor images across groups adjusting for other covariates, which is of central interest in neuroimaging analysis. Second, while existing regularization solutions largely rely on penalty functions, our envelope based method provides an alternative way of introducing regularization into estimation. It complements the usual penalty function based solutions, and usefully expands the realm of regularized estimation in general. Moreover, our method can be naturally coupled with penalty functions for further regularization. Third, our work advances the recent development of envelope method that was first proposed by \citet{Cook2010} then further developed in a series of papers \citep{SuCook2011, SuCook2012, SuCook2013, Cook2013PLS, CFZ2014, CookZhang2014simultaneous, CookZhang2014foundation}. Whilst all existing envelope methods concentrate on a scalar or vector response, our work differs obviously by tackling a tensor response. Such an extension is far from trivial, and new techniques are required throughout its development, even though to make our proposal easier to comprehend, we have chosen to present our method in a way that is parallel to that for a vector response. Furthermore, since the envelope methodology is new and sometimes uneasy to follow, we strive to connect it with the more familiar sparsity principle and clearly outline its assumptions, gains and limitations.

The rest of the article is organized as follows. Section~\ref{sec:preparation} reviews key tensor notations and operations, and summarizes the envelope method for multivariate vector response regression. Section~\ref{sec:model} proposes tensor response linear model, the generalized sparsity principle, then the concept of tensor envelope. Section~\ref{sec:estimation} develops two estimators, and Section~\ref{sec:theory} studies their asymptotic properties. Simulations and real data analyses are presented in Sections~\ref{sec:simulation} and \ref{sec:realdata}, followed by a discussion in Section~\ref{sec:discussion}. All technical proofs are relegated to the Supplementary Materials.

\section{Preparations}
\label{sec:preparation}

\subsection{Tensor notations and operations}

We begin with a quick review of some tensor notations and operations that are to be intensively used in this article. See also \citet{KoldaBader09Tensor} for an excellent review.

Multidimensional array $\Abf\in \real{r_{1}\times\cdots\times r_{m}}$ is called an \emph{$m$th-order tensor}. The order of a tensor is also known as \emph{dimension}, \emph{way} or \emph{mode}. A \emph{fiber} is the higher order analogue of matrix row and column, and is defined by fixing every index of the tensor but one. A matrix column is a mode-1 fiber and a row is a mode-2 fiber.

The \emph{$\vect(\Abf)$ operator} stacks the entries of a tensor into a column vector, so that an entry $a_{i_1\ldots i_m}$ of $\Abf$ maps to the $j$-th entry of $\vect(\Abf)$, in which $j = 1 + \sum_{k=1}^m (i_k-1) \prod_{k'=1}^{k-1} r_{k'}$.
The \emph{mode-$k$ matricization}, $\Abf_{(k)}$, maps a tensor $\Abf$ into a matrix, denoted by $\Abf_{(k)}\in\real{r_{k}\times(\prod_{j\neq k}r_{j})}$, so that the $(i_1,\ldots,i_m)$ element of $\Abf$ maps to the $(i_k,j)$ element of the matrix $\Abf_{(k)}$, where $j = 1 + \sum_{k'\ne k} (i_{k'}-1) \prod_{k''<k',k'' \ne k} r_{k''}$. The \emph{$k$-mode product} of a tensor $\Abf$ and a matrix $\Cbf\in\real{s\times r_{k}}$ results in an $m$th-order tensor denoted as $\Abf\times_{k} \Cbf \in \real{r_{1}\times\cdots\times r_{k-1}\times s\times r_{k+1}\times\cdots\times r_m}$, where each element is the product of mode-$k$ fiber of $\Abf$ multiplied by $\Cbf$. Similarly, the \emph{$k$-mode vector product} of a tensor $\Abf$ and a vector $\cbf\in\real{r_k}$ results in an $(m-1)$th-order tensor denoted as $\Abf \bar{\times}_{k} \; \cbf \in \real{r_{1}\times\cdots\times r_{k-1}\times r_{k+1}\times\cdots\times r_m}$, where each element is the inner product of each mode-$k$ fiber of $\Abf$ with the vector $\cbf$.

The \emph{Tucker decomposition} of a tensor is defined as $\Abf = \Cbf\times_{1}\Gammabf^{(1)}\times_{2}\cdots\times_{m}\Gammabf^{(m)}$, where $\Cbf\in\real{u_{1}\times\cdots\times u_{m}}$ is the \emph{core tensor}, and $\Gammabf^{(k)}\in\real{r_{k}\times u_{k}}$, $k=1,\dots,m$, are the factor matrices. It is a low-rank decomposition of the original tensor $\Abf$. For convenience, the Tucker decomposition is often represented by a shorthand, $\llbracket\Cbf;\Gammabf^{(1)},\dots,\Gammabf^{(m)}\rrbracket$.

\subsection{Multivariate response envelope model}
\label{sec:vector-resp}

Next we briefly review the multivariate linear model with vector-valued response, along with some key concepts of envelope, and with two goals in mind. First, it is to assist with a better understanding of the envelope methods in general, and second, to facilitate the construction of envelopes for tensor-valued response regression.

We start with the classical multivariate response linear model,
\begin{eqnarray} \label{eqn:model-vector}
\Ybf = \betabf \Xbf + \varepsilonbf,
\end{eqnarray}
where $\Ybf \in \real{r}$ is a response vector, $\Xbf \in \real{p}$ is a predictor vector, $\betabf \in \real{r \times p}$ is the coefficient matrix, while the intercept is set to zero by centering the samples of $\Ybf$ and $\Xbf$, and $\varepsilonbf \in \real{r}$ is the i.i.d. error that is independent of $\Xbf$. It is often assumed that $\varepsilonbf$ follows a multivariate normal distribution with mean zero and covariance $\Sigmabf \in \real{r \times r}$, $\Sigmabf>0$, though normality is not essential.

The envelope model \citep{Cook2010} builds upon a key assumption that some aspects of the response vector are stochastically constant as the predictors vary. In other words, part of the response is irrelevant to the regression. More rigorously, we assume there exists a subspace $\spc$ of $\real{r}$ such that
\begin{eqnarray} \label{eqn:assumption-vector}
\Qbf_\spc \Ybf | \Xbf \sim \Qbf_\spc \Ybf, \quad \quad \Qbf_\spc \Ybf \indep \Pbf_\spc \Ybf | \Xbf,
\end{eqnarray}
where $\Pbf_\spc$ is the projection matrix onto $\spc$, $\Qbf_\spc = \Ibf_r - \Pbf_{\spc}$ is the projection onto the complement space $\cspc$, $\sim$ means identically distributed, and $\indep$ means statistical independence. To better understand this assumption, we introduce a basis system of $\spc$. Let $\Gammabf \in \real{r \times u}$ denote a basis matrix of $\spc$, where $u$ is the dimension of $\spc$, $u \leq r$, and $\Gammabf_0 \in \real{r \times (r-u)}$ a basis of $\cspc$. Then \eqref{eqn:assumption-vector} essentially states that the linear combinations $\Gammabf_0\trans \Ybf \in \real{r-u}$ are immaterial to the estimation of $\betabf$ in that it responds neither to changes in the predictors nor to those in the rest of the response $\Gammabf\trans \Ybf \in \real{u}$. Correspondingly, $\Gammabf\trans \Ybf$ carry all the material information in $\Ybf$, and intuitively, one can focus on $\Gammabf\trans \Ybf$ in regression modeling.

We remark that, assumption \eqref{eqn:assumption-vector}, although looks somewhat unfamiliar, can find its natural counterpart in the well known and commonly adopted \emph{sparsity principle} in classical variable selection, where only a subset of variables are assumed to be relevant to regressions. The two assumptions, at heart, share exactly the same spirit that only part of information is deemed useful for regressions and the rest irrelevant. However, they are also different in that, whereas the usual sparsity principle focuses on \emph{individual} variables, \eqref{eqn:assumption-vector} permits \emph{linear combination} of the variables to be irrelevant. For this reason, we term assumption \eqref{eqn:assumption-vector} as the \emph{generalized sparsity principle}. Compared to the usual sparsity principle, it is more flexible, but could lose some interpretability.

To see how the generalized sparsity principle would facilitate estimation of $\betabf$ in model \eqref{eqn:model-vector}, we note that the following decompositions hold true under \eqref{eqn:assumption-vector}.
\begin{eqnarray*}
\spn(\betabf) \subseteq \spc \;\; \mathrm{ and } \;\; \Sigmabf = \var(\Pbf_\spc \Ybf) + \var(\Qbf_\spc \Ybf) = \Pbf_\spc \Sigmabf \Pbf_\spc + \Qbf_\spc \Sigmabf \Qbf_\spc,
\end{eqnarray*}
where $\spn(\betabf)$ is the column subspace of $\betabf$, i.e. the subspace spanned by the columns of $\betabf$.
Accordingly, we can rewrite the above decompositions in terms of the basis matrices,
\begin{eqnarray} \label{eqn:decomp-vector}
\betabf = \Gammabf\thetabf \;\; \mathrm{ and } \;\; \Sigmabf = \Gammabf \Omegabf \Gammabf\trans + \Gammabf_{0} \Omegabf_{0} \Gammabf_{0}\trans,
\end{eqnarray}
where $\thetabf=\Gammabf\trans \betabf \in \real{u \times p}$ denotes the coordinates of $\betabf$ relative to the basis $\Gammabf$, $\Omegabf = \cov(\Gammabf\trans\Ybf\mid \Xbf) \in \real{u \times u}$ is the material variation, and $\Omegabf_{0}=\cov(\Gammabf\trans_0\Ybf\mid \Xbf) = \cov(\Gammabf\trans_0\Ybf) \in \real{(r-u) \times (r-u)}$ is the immaterial variation that contains no information about $\Ybf | \Xbf$, but only brings extraneous variation in estimation.

Given the first result of \eqref{eqn:decomp-vector}, we note that model \eqref{eqn:model-vector} can be rewritten as
\begin{eqnarray} \label{eqn:reduced-resp}
\Gammabf\trans \Ybf = \thetabf \Xbf + \Gammabf\trans \varepsilonbf, \;\; \mathrm{ and } \;\;
\Gammabf_0\trans \Ybf = \Gammabf_0\trans \varepsilonbf.
\end{eqnarray}
In turn \eqref{eqn:reduced-resp} implies that regression modeling can now focus on the material part $\Gammabf\trans \Ybf$ only. The effective number of parameters in model \eqref{eqn:model-vector} is reduced from $pr + r(r+1)/2$ without assumption \eqref{eqn:assumption-vector}, to $pu + (r-u)u + u(u+1)/2 + (r-u)(r-u+1)/2$ with the assumption, and the difference is $p(r-u)$.

Given the second result of \eqref{eqn:decomp-vector}, \citet{Cook2010} showed the gain in estimation efficiency for $\betabf$. Let
$\bhatenv$ denote the estimator of $\betabf$ in \eqref{eqn:model-vector} under \eqref{eqn:assumption-vector}, $\bhatols$ the ordinary least squares estimator without imposing assumption \eqref{eqn:assumption-vector}, and $\betabf_{\textrm{{\tiny TRUE}}}$ the true value of $\betabf$. Then it is shown that, both $\sqrt{n} \left\{ \vect(\bhatols) - \vect(\betabf_{\textrm{{\tiny TRUE}}}) \right\}$ and $\sqrt{n} \left\{ \vect(\bhatenv) - \vect(\betabf_{\textrm{{\tiny TRUE}}}) \right\}$ converge to a normal vector with mean zero and covariance matrix,
\begin{eqnarray*}
\avar\left\{\sqrt{n} \; \vect(\bhatols) \right\} = \Sigx^{-1} \otimes \Sigmabf, \; \textrm{ and } \avar\left\{\sqrt{n} \; \vect(\bhatenv) \right\} = \Sigx^{-1} \otimes \left( \Gammabf \Omegabf \Gammabf\trans \right) + \Deltabf,
\end{eqnarray*}
respectively, where $\Sigx = \cov(\Xbf)$, the first term in $\avar \{\sqrt{n} \; \vect(\bhatenv)\}$ corresponds to the asymptotic variance of the estimator given that $\spc$ is known, and the second term $\Deltabf$ is the asymptotic cost of estimating $\spc$. While \citet{Cook2010} showed that $\avar \{\sqrt{n} \; \vect(\bhatenv)\} \leq \avar \{\sqrt{n} \; \vect(\bhatols)\}$ in general, in the light of decomposition of $\Sigmabf$ in \eqref{eqn:decomp-vector},  it is straightforward to see that the first term in $\avar \{\sqrt{n} \; \vect(\bhatenv)\}$ is to be substantially smaller than $\avar \{\sqrt{n} \; \vect(\bhatols)\}$, if the immaterial variation $\Gammabf_{0} \Omegabf_{0} \Gammabf_{0}\trans$ dominates the material variation $\Gammabf \Omegabf \Gammabf\trans$.

Finally, we address the issue of existence and uniqueness of $\spc$ in \eqref{eqn:assumption-vector}. The subspace $\spc$ always exists, as it can trivially take the form of $\real{r}$. However, $\spc$ is not unique. Then the idea is to seek the \emph{intersection} of all subspaces that satisfy \eqref{eqn:assumption-vector}, which is minimum and unique. Toward that end, \citet{Cook2010} gave two generic definitions.

\begin{definition} \label{def:reducespace} A subspace ${\mathcal R} \subseteq \real{p}$ is said to be a reducing subspace of $\Mbf \in \real{p\times p}$ if ${\mathcal R}$ satisfies that $\Mbf = \Pbf_{{\mathcal R}} \Mbf \Pbf_{{\mathcal R}} + \Qbf_{{\mathcal R}} \Mbf\Qbf_{{\mathcal R}}$.
\end{definition}

\begin{definition}\label{def:envelope-generic} Let $\Mbf \in \real{p\times p}$ and ${\mathcal B} \subseteq \spn(\Mbf)$. Then the $\Mbf$-envelope of ${\mathcal B}$, denoted by $\env_{\Mbf}({\mathcal B})$, is the intersection of all reducing subspaces of $\Mbf$ that contain ${\mathcal B}$.
\end{definition}

\noindent
Given those definitions, we see that any subspace $\spc$ satisfying \eqref{eqn:assumption-vector} under model \eqref{eqn:model-vector} is a reducing subspace of $\Sigmabf$, and the intersection of all such subspaces is the $\Sigmabf$-envelope of ${\mathcal B} = \spn(\betabf)$. This envelope $\env_{\mathbf{\Sigma}}(\mathcal{B})$ is also denoted by $\env_{\mathbf{\Sigma}}(\betabf)$, and uniquely exists. In our envelope based estimation, we seek the estimation of $\env_{\mathbf{\Sigma}}(\betabf)$ so to improve estimation of the coefficient matrix $\betabf$.

\section{Models}
\label{sec:model}

\subsection{Tensor response linear model}

When facing a tensor response, we develop a model in analogous to the classical multivariate model \eqref{eqn:model-vector}. That is, for an $m$th order tensor response $\Ybf \in \real{r_1 \times \cdots \times r_m}$, and a vector of predictors $\Xbf \in \real{p}$, consider the tensor response linear model
\begin{equation} \label{eqn:model-tensor}
\Ybf = \Bbf \bar{\times}_{(m+1)} \Xbf + \varepsilonbf.
\end{equation}
In this model, the coefficient $\Bbf \in \real{r_1 \times \cdots \times r_m \times p}$ is an $(m+1)$th order tensor that captures the interrelationship between $\Ybf$ and $\Xbf$ and is the parameter we are primarily interested in. $\bar{\times}_{(m+1)}$ is the $(m+1)$-mode vector product. The intercept term is again omitted without losing any generality. The error $\varepsilonbf \in \real{r_1 \times \cdots\times r_m}$ is an $m$th order tensor that is independent of $\Xbf$ and has mean zero. Furthermore, we assume that $\varepsilonbf$ has a \emph{separable Kronecker covariance structure} such that $\cov\{\vect(\varepsilonbf)\} = \Sigmabf = \Sigmabf_{m} \otimes \cdots \otimes \Sigmabf_{1}$, $\Sigmabf_k > 0, k = 1,\ldots,m$. This separable covariance assumption is important to help reduce the number of free parameters in $\Sigmabf$, which is part of our envelope estimation. Meanwhile, this separable structure has been fairly commonly imposed in the tensor literature; see for instance, \citet{Hoff2011, FosdickHoff2014}. Here to avoid notation proliferation, we continue to use $\Sigmabf$ to denote the covariance matrix, as it should not cause any confusion in the context. The distribution of $\vect(\varepsilonbf)$ is assumed to be normal, which enables likelihood estimation. However, normality is not essential, and moment based estimation can replace likelihood estimation when the normality assumption is in question.

Two special cases of model \eqref{eqn:model-tensor} are worth of brief mentioning. The first is when $\Xbf$ is a scalar and takes the value of only $0$ or $1$. In this case, $\Bbf$ reduces to an $m$th order tensor that can be interpreted as the mean difference of the tensor coefficients between the two populations. The second case is when $m=1$, where the response $\Ybf$ becomes a vector, and model \eqref{eqn:model-tensor} reduces to the classical multivariate linear model \eqref{eqn:model-vector}. In this case, $\Bbf \bar{\times}_{(m+1)} \Xbf$ becomes the inner product of each mode-$2$ fiber (i.e., row) of $\Bbf$ with $\Xbf$, which in turn is the usual matrix product of $\Bbf$ and $\Xbf$.

Next we consider an alternative tensor response linear model \eqref{eqn:model-tensor},
\begin{eqnarray} \label{eqn:model-tvec}
\vect(\Ybf) = \Bbf_{(m+1)}\trans \Xbf + \vect(\varepsilonbf).
\end{eqnarray}
This model can be viewed as the \emph{vectorized} form of model \eqref{eqn:model-tensor}. However, the main difference is that \emph{no} separable covariance structure is imposed on the error term $\varepsilonbf$ in \eqref{eqn:model-tvec}. The coefficient matrix $\Bbf_{(m+1)} \in \real{p \times \prod_{k=1}^{m} r_k}$ can be interpreted as the mode-$(m+1)$ matricization of the tensor coefficient $\Bbf$ in \eqref{eqn:model-tensor}. Each column of $\Bbf_{(m+1)}$ is a $p \times 1$ coefficient vector that characterizes the linear relationship between each individual element of $\Ybf$ and the predictor vector $\Xbf$. Therefore, estimating $\Bbf_{(m+1)}$ in \eqref{eqn:model-tvec} is equivalent to estimating $\Bbf$ in \eqref{eqn:model-tensor} by fitting individual elements of $\Ybf$ on $\Xbf$ \emph{one at a time}. We call this estimator the ordinary least squares estimator of $\Bbf$, and denote it by $\Bhatols$. Given the data $\{ (\Xbf_i, \Ybf_i) \}_{i=1}^{n}$, it has an explicit form
\begin{eqnarray*}\label{eqn:tensorOLS}
\Bhatols = \mathbb{Y} \times_{(m+1)} \{ (\mathbb{X} \mathbb{X}\trans)^{-1} \mathbb{X} \},
\end{eqnarray*}
where $\mathbb{X} \in \real{p \times n}$ and $\mathbb{Y} \in \real{r_1 \times \ldots \times r_m \times n}$ are the \emph{stacked} predictor matrix and response array, respectively. If $\vect(\varepsilonbf)$ is further assumed to be normally distributed, then the above OLS estimator is also the maximum likelihood estimator based on model \eqref{eqn:model-tvec}.

\subsection{Generalized sparsity principle}

For a tensor response, we expect a similar generalized sparsity principle like \eqref{eqn:assumption-vector} to hold true. It is probably more so than the vector response case, as intuitively it is reasonable to expect certain regions of the tensor response to be immaterial. More specifically, suppose there exist a series of subspaces, $\spc_k\subseteq\real{r_k}$, $k=1,\ldots,m$, such that
\begin{eqnarray} \label{eqn:assumption-tensor}
\Ybf \times_k \Qbf_k | \Xbf \sim \Ybf \times_k \Qbf_k, \quad \Ybf \times_k \Qbf_k \indep \Ybf \times_k \Pbf_k | \Xbf, \quad k = 1,\ldots,m,
\end{eqnarray}
where $\Pbf_k \in \real{r_k \times r_k}$ is the projection matrix onto $\spc_k$, $\Qbf_k = \Ibf_{r_k} - \Pbf_k \in \real{r_k \times r_k}$ is the projection onto the complement space $\cspc_k$, and $\times_k$ denotes the $k$-mode product. Then, the first condition in \eqref{eqn:assumption-tensor} essentially states that $\Ybf \times_k \Qbf_k$ does not depend on $\Xbf$, while the second condition in \eqref{eqn:assumption-tensor} says $\Ybf \times_k \Qbf_k$ does not affect the rest of the response, $\Ybf \times_k \Pbf_k$, and there is no information leak between $\Ybf \times_k \Qbf_k$ and $\Ybf \times_k \Pbf_k$. As such, we call $\Ybf \times_k \Qbf_k$ the immaterial information to the regression of $\Ybf$ on $\Xbf$, and call $\Ybf \times_k \Pbf_k$ the material information. Combining the statements in \eqref{eqn:assumption-tensor} for all $k = 1,\ldots, m$, we arrive at a parsimonious representation:
\begin{equation*}\label{eqn:assumption-tensor-all}
\Qtensor(\Ybf) | \Xbf \sim \Qtensor(\Ybf), \quad \quad \Qtensor(\Ybf) \indep \Ptensor(\Ybf) | \Xbf,
\end{equation*}
where $\Qtensor(\Ybf) = \Ybf - \Ptensor(\Ybf) \in \real{r_1\times \ldots \times r_m}$, and $\Ptensor(\Ybf) = \llbracket  \Ybf; \Pbf_{1},\dots,\Pbf_{m} \rrbracket \in \real{r_1\times \ldots \times r_m}$, i.e., a Tucker decomposition with $\Ybf$ as the core tensor, and $\Pbf_{1},\dots,\Pbf_{m}$ as the factor matrices along each mode. These two operators provide a decomposition of $\Ybf$, $\Ybf = \Ptensor(\Ybf) + \Qtensor(\Ybf)$, into the material part $\Ptensor(\Ybf)$ and the immaterial part $\Qtensor(\Ybf)$.

Introducing \eqref{eqn:assumption-tensor} to the tensor response linear model \eqref{eqn:model-tensor}, we have the following results.
\begin{proposition}\label{thm:equivalence}
Under the tensor response linear model \eqref{eqn:model-tensor}, the assumption \eqref{eqn:assumption-tensor} is true if and only if
\begin{equation*}
\Bbf\times_{k} \Qbf_{k} = 0 \;\; \textrm{ and } \;\; \Sigmabf_{k} = \Pbf_{k} \Sigmabf_{k} \Pbf_{k} + \Qbf_{k} \Sigmabf_{k} \Qbf_{k}, \;\;\; k=1,\dots,m.
\end{equation*}
\end{proposition}
\noindent
To turn the above decompositions of $\Bbf$ and $\Sigmabf_{k}$ into a basis representation, let $\Gammabf_k \in \real{r_k \times u_k}$ denote a basis  for $\spc_{k}$, where $u_k$ is the dimension of $\spc_{k}$, and $\Gammabf_{0k} \in \real{r_k \times (r_k - u_k)}$ denote the complement basis, $k = 1, \ldots, m$. Let $\Omegabf_{k} \in \real{u_k \times u_k}$ and $\Omegabf_{0k} \in \real{(r_k - u_k) \times (r_k - u_k)}$ denote two symmetric positive definite matrices. Then we have $\Pbf_k=\Gammabf_k\Gammabf_k\trans$, $\Qbf_k=\Gammabf_{0k}\Gammabf_{0k}\trans$, plus the following parameterization for $\Bbf$ and $\Sigmabf = \Sigmabf_{m} \otimes \cdots \otimes \Sigmabf_{1}$.

\begin{proposition}\label{thm:parameter} The parameterization in Proposition~\ref{thm:equivalence} is equivalent to the following coordinate representations:
\begin{eqnarray*}
\Bbf & = & \llbracket \Thetabf; \Gammabf_1, \ldots, \Gammabf_m, \Ibf_p \rrbracket \;\; \textrm{ for some } \Thetabf \in \real{u_1 \times \ldots \times u_m \times p}, \\
\Sigmabf_k & = & \Gammabf_k \Omegabf_k \Gammabf_k\trans + \Gammabf_{0k} \Omegabf_{0k} \Gammabf_{0k}\trans, \;\; k = 1, \ldots, m.
\end{eqnarray*}
\end{proposition}
\noindent
Accordingly, one can rewrite the material response part $\Ptensor(\Ybf)$ in the following way
\begin{equation*}\label{eqn:Ptensor}
\Ptensor(\Ybf) = \llbracket  \Ybf; \Gammabf_{1}\Gammabf_1\trans,\dots,\Gammabf_{m}\Gammabf_m\trans \rrbracket
=\llbracket \llbracket \Ybf;\Gammabf\trans_1,\dots,\Gammabf\trans_m \rrbracket; \Gammabf_1,\dots,\Gammabf_m \rrbracket,
\end{equation*}
where the \emph{core tensor} is $\Zbf = \llbracket \Ybf;\Gammabf_1\trans,\dots,\Gammabf\trans_m \rrbracket \in \real{u_1\times \ldots \times u_m}$. We see that, by recognizing and focusing on the material part of the tensor response $\Ptensor(\Ybf)$, the regression modeling can now focus on the core tensor $\Zbf$ as the ``surrogate response", which plays a similar role as $\Gammabf\trans \Ybf$ in \eqref{eqn:reduced-resp} in the vector response case. Meanwhile, the key parameter to estimate becomes $\Thetabf \in \real{u_1 \times \ldots \times u_m \times p}$, along with $\{ \Gammabf_k \}_{k=1}^{m}$, $\{ \Omegabf_{k} \}_{k=1}^{m}$ and $\{ \Omegabf_{0k} \}_{k=1}^{m}$. Consequently, the number for free parameters reduces from $p \prod_{k=1}^{m}r_{k} + \sum_{k=1}^{m}r_{k}(r_{k}+1)/2$ to $p \prod_{k=1}^{m}u_{k} + \sum_{k=1}^{m}\{u_{k}(r_{k}-u_{k}) + u_{k}(u_{k}+1)/2 + (r_{k}-u_{k})(r_{k}-u_{k}+1)/2\}$, and in effect saving $p \bigl\{ \prod_{k=1}^{m} r_{k} - \prod_{k=1}^{m} u_{k} \bigr\}$ parameters. With $u_k$ usually being much smaller than $r_k$, substantial dimension reduction is achieved, which in turn improves the estimation.

\subsection{Tensor envelope}

Similar to the vector case, we next develop the notion of \emph{tensor envelope} for tensor response model \eqref{eqn:model-tensor} to attain uniqueness of the subspaces $\spc_k$ in the generalized sparsity principle \eqref{eqn:assumption-tensor}. Unlike the vector case, however, there are two different ways to construct a tensor envelope. We will define the new concept in one way, then establish its equivalence with the other. Moreover, we will lay out the difference between the proposed tensor envelope and the vector envelope that is constructed based on the vectorized model \eqref{eqn:model-tvec}.

One way to establish the tensor envelope for model \eqref{eqn:model-tensor} is to recognize that it should contain $\spn(\Bbf_{(m+1)}\trans)$, meanwhile it should reduce the covariance $\Sigmabf$ and respect the separable Kronecker covariance structure that $\Sigmabf = \Sigmabf_{m} \otimes \cdots \otimes \Sigmabf_{1}$. Then following Definitions~\ref{def:reducespace} and \ref{def:envelope-generic}, we come to the next definition of the tensor envelope.

\begin{definition} \label{def:envelope-tensor} The tensor envelope for $\Bbf$ in the tensor response linear model \eqref{eqn:model-tensor}, denoted by $\envt$, is defined as the intersection of all reducing subspaces $\env$ of $\Sigmabf$ that contains $\spn(\Bbf_{(m+1)}\trans)$ and can be written as $\env = \env_{m} \otimes \cdots \otimes \env_{1}$, where $\env_{k} \subseteq \real{r_{k}}$, $k=1,\dots,m$.
\end{definition}

\noindent
Following this definition, we see that $\envt$ is minimum and unique, and is of central interest in our envelope based estimation of $\Bbf$. Moreover, under the special case that $m=1$ and the response is a vector, the tensor envelope $\envt$ reduces to the envelope notion $\envs$ for the vector response.

An alternative way to construct the tensor envelope is by noting that, due to the decomposition in Proposition~\ref{thm:equivalence}, one can construct a series of envelopes, $\envsk{k}$, for $k=1,\ldots,m$, that satisfy the generalized sparsity principle \eqref{eqn:assumption-tensor} under model \eqref{eqn:model-tensor}. That is, $\envsk{k}$ is the smallest subspace $\spc_k$ that contains $\spn(\Bbf_{(k)})$ and reduces $\Sigmabf_k$, $k=1,\ldots,m$. Then one can construct a tensor envelope by combining $\left\{ \envsk{k} \right\}_{k=1}^{m}$ to capture all the material information in the response. Naturally, the two ways of constructing the tensor envelope are well connected, due to the next equivalence property.

\begin{proposition}\label{thm:tenv-eq} The tensor envelope as defined in Definition~\ref{def:envelope-tensor} satisfies that $\envt = \envsk{m} \otimes \cdots \otimes \envsk{1}$.
\end{proposition}
\noindent
Our estimation of the tensor envelope utilizes this result by first estimating a basis of the individual envelope $\envsk{k}$, then combining them by Kronekcer product to construct an estimate of the tensor envelope.

Finally, we remark that, in principle, one can develop an envelope for the ordinary least squares estimator in model \eqref{eqn:model-tvec} as well. By analogy to the envelope definition for a vector response, this envelope also contains $\spn(\Bbf_{(m+1)}\trans)$, and thus we denote it by ${\mathcal E}_{\bm\Sigma}(\Bbf_{(m+1)})$. However, there are some important differences between ${\mathcal E}_{\bm\Sigma}(\Bbf_{(m+1)})$ and the tensor envelope $\envt$ in Definition~\ref{def:envelope-tensor}. First, ${\mathcal E}_{\bm\Sigma}(\Bbf_{(m+1)})$ does not take into account the separable covariance structure, nor can be decomposed into the Kronecker product of the individual envelopes $\envsk{k}$. Second, the computation involved in estimating ${\mathcal E}_{\bm\Sigma}(\Bbf_{(m+1)})$ is prohibitive, as it replies on the estimation of the unstructured covariance matrix of $\vect(\varepsilonbf) \in \real{\prod_{k=1}^{m} r_k \times \prod_{k=1}^{m} r_k}$. By contrast, the computation of $\envt$ is much more feasible, as we demonstrate in the next section.

\section{Estimation}
\label{sec:estimation}

Our primary target of estimation is $\Bbf$ in the tensor response linear model \eqref{eqn:model-tensor}. Our proposal is to estimate $\Bbf$ through the tensor envelope $\envt$, which also involves estimation of $\Sigmabf = \Sigmabf_{m} \otimes \cdots \otimes \Sigmabf_{1}$. The objective function to minimize is of the form,
\begin{equation*}\label{eqn:ellBSigma}
\ell(\Bbf, \Sigmabf) = \log| \Sigmabf |+n^{-1}\sum_{i=1}^{n} \left\{ \vect(\Ybf_{i}) - \Bbf_{(m+1)}\trans \Xbf_{i} \right\}\trans \Sigmabf^{-1} \left\{ \vect(\Ybf_{i}) - \Bbf_{(m+1)}\trans \Xbf_{i} \right\}.
\end{equation*}
It is straightforward to verify that this objective function, aside from some constant, is the negative log-likelihood function of the model \eqref{eqn:model-tensor} if one assumes that the error follows a normal distribution. By adopting \eqref{eqn:assumption-tensor} then the parameter decompositions in Proposition~\ref{thm:parameter}, the minimization of $\ell(\Bbf, \Sigmabf)$ becomes estimation of the envelope basis $\Gammabf_k \in \real{r_k \times u_k}$, the reduced coefficient $\Thetabf \in \real{u_1 \times \ldots \times u_m \times p}$, and the matrices $\Omegabf_{k} \in \real{u_k \times u_k}$ and $\Omegabf_{0k} \in \real{(r_k - u_k) \times (r_k - u_k)}$, $k = 1, \ldots, m$. Here, with a slight abuse of notation, we continue to denote the dimension of the individual envelope $\envsk{k}$ as $u_k$.

We present two solutions, one an iterative estimator and the other a one-step estimator. The first solution alternates among steps of estimating one parameter given the rest fixed. It leads to a maximum likelihood estimator when the error follows a normal distribution and is a moment estimator otherwise. The second solution requires no iteration, and is essentially an approximate estimator, but it enjoys several appealing properties, both computationally and theoretically.

\subsection{Iterative estimator}
\label{sec:iterative}

We first summarize our iterative optimization of $\ell(\Bbf, \Sigmabf)$ in Algorithm~\ref{algo:all}. We then give details for each individual step. Updating equations in each step are carefully derived as partial maximized likelihood estimators under the normal error assumption, with the detailed derivation given in the Supplementary Materials. As a result, the objective function is monotonically decreasing, guaranteeing the convergence of the algorithm.

\begin{algorithm}[t]
\begin{algorithmic}
\STATE [1] Initialize $\Bbf^{(0)}$ and $\Sigmabf^{(0)} = \Sigmabf_{m}^{(0)} \otimes \cdots \otimes \Sigmabf_{1}^{(0)}$
\REPEAT
\STATE [2] Estimate envelope basis $\bigl\{ \Gammabf_k^{(t+1)} \bigr\}_{k=1}^m$ given $\Bbf^{(t)}$ and $\Sigmabf^{(t)}$
\STATE [3] Estimate parameters $\Thetabf^{(t+1)}$,  $\Omegabf_k^{(t+1)}$ and $\Omegabf_{0k}^{(t+1)}$ given $\bigl\{ \Gammabf_k^{(t+1)} \bigr\}_{k=1}^m$.
\STATE [4] Update $\Bbf^{(t+1)}$ and $\Sigmabf^{(t+1)} = \Sigmabf_{m}^{(t+1)} \otimes \cdots \otimes \Sigmabf_{1}^{(t+1)}$.
\UNTIL{the objective function converges}
\end{algorithmic}
\caption{\label{algo:all} Iterative optimization algorithm for minimizing $\ell(\Bbf, \Sigmabf)$.}
\end{algorithm}

The first step of Algorithm~\ref{algo:all} is to initialize $\Bbf$ and $\Sigmabf$. For $\Bbf$, a natural initial estimator is the OLS estimator $\Bhatols$ in \eqref{eqn:tensorOLS}. That is, we fit each element of the tensor response $\Ybf$ on $\Xbf$ one at a time, and set the resulting estimator as the initial estimator $\Bbf^{(0)} = \Bhatols$. For $\Sigmabf$, we employ the covariance estimator of \citet{Dutilleul1999} and \citet{ManceurDutilleul2013}. That is, for $k = 1, \ldots, m$, in turn, we set
\begin{eqnarray*}\label{eqn:covMLE}
\Sigmabf_{k}^{(0)} = \frac{1}{n \prod_{j\neq k} r_{j}} \sum_{i=1}^{n} \ebf_{i(k)} \left\{ (\Sigmabf_{m}^{(0)})^{-1} \otimes \ldots \otimes (\Sigmabf_{k+1}^{(0)})^{-1} \otimes (\Sigmabf_{k-1}^{(0)})^{-1} \otimes \ldots \otimes (\Sigmabf_{1}^{(0)})^{-1} \right\} \ebf_{i(k)}\trans,
\end{eqnarray*}
where $\ebf_{i(k)}$ is the mode-$k$ matricization of the residual, $\ebf_i = \Ybf_i - \Bbf^{(0)} \times_{(m+1)} \Xbf_i, i = 1, \ldots, n$, for $n$ sample replications, and the iterative update of each covariance $\Sigmabf_{k}^{(0)}$ given the rest starts with $\Sigmabf_{j}^{(0)}=\Ibf_{r_j}$, $j\neq k$. One can verify that the above estimator is the maximum likelihood estimator if the error in \eqref{eqn:model-tensor} follows a normal distribution. We also remark that, the individual component $\Sigmabf_k$ is not identifiable. To address this issue, we normalize $\Sigmabf_k^{(0)}$ by dividing the term by its Frobenius norm, and update the final estimator $\Sigmabf^{(0)} = \tau \Sigmabf_m^{(0)} \otimes \ldots \otimes \Sigmabf_1^{(0)}$, where the scalar $\tau = (n\prod_j r_j)^{-1}\sum_{i=1}^n \vect\trans(\ebf_i) \bigl\{ (\Sigmabf_{m}^{(0)})^{-1}\otimes\cdots\otimes(\Sigmabf_{1}^{(0)})^{-1} \bigr\}
\vect(\ebf_i)$.

The second step of Algorithm~\ref{algo:all} is to estimate the envelope basis $\{ \Gammabf_k \}_{k=1}^{m}$. Update of $\Gammabf_k^{(t+1)}$, given the current estimates $\Bbf^{(t)}$ and $\Sigmabf^{(t)}$, can be achieved by minimizing the following objective function, subject to the constraint that $\Gbf_k\trans\Gbf_k=\Ibf_{u_k}$,
\begin{equation}\label{eqn:lik_obj_Gk}
f_k^{(t)}(\Gbf_k) = \log\vert\Gbf\trans_k\Mbf_k^{(t)}\Gbf_k\vert + \log\vert\Gbf\trans_k (\Nbf_{k}^{(t)})^{-1}\Gbf_k\vert,
\end{equation}
where $\Mbf^{(t)}_k=(n \prod_{j\neq k} r_{j})^{-1} \sum_{i=1}^{n}\deltabf^{(t)}_{i(k)}\{(\Sigmabf_m^{(t)})^{-1}\otimes\cdots\otimes
(\Sigmabf_{k+1}^{(t)})^{-1}\otimes(\Sigmabf_{k-1}^{(t)})^{-1}\otimes\cdots\otimes(\Sigmabf_1^{(t)})^{-1}\} (\deltabf_{i(k)}^{(t)})\trans$, $\Nbf_{k}^{(t)}=(n \prod_{j\neq k} r_{j})^{-1} \sum_{i=1}^{n}\Ybf_{i(k)}\{(\Sigmabf_m^{(t)})^{-1}\otimes\cdots\otimes (\Sigmabf_{k+1}^{(t)})^{-1}\otimes(\Sigmabf_{k-1}^{(t)})^{-1}\otimes\cdots\otimes(\Sigmabf_1^{(t)})^{-1}\} \Ybf_{i(k)}\trans$, and $\deltabf_{i}^{(t)}$ is the fitted envelope model residual were the envelope basis $\{\Gammabf_j\}_{j=1, j\neq k}^{m}$ known:
%\begin{equation}\label{eqn:deltai}
$\deltabf_i^{(t)}  =  \Ybf_i - \llbracket \Bhatols; \PGammak{1}^{(t)}, \ldots, \PGammak{k-1}^{(t)}, \Ibf_{r_k}, \PGammak{k+1}^{(t)}, \ldots, \PGammak{m}^{(t)}, \Ibf_p \rrbracket\times_{(m+1)}\Xbf_i\trans$,
%\end{equation}
where $\PGammak{j}^{(t)}\equiv\Gammabf^{(t)}_{j} (\Gammabf^{(t)}_{j})\trans$ is the projection onto subspace $\spn(\Gammabf^{(t)}_{j})$ and $\deltabf^{(t)}_{i(j)}$ is the mode-$j$ matricization of $\deltabf^{(t)}_{i}$. Then we have $\Gammabf_k^{(t+1)} = \arg\min_{\Gbf_k} f_k^{(t)}(\Gbf_k)$ subject to the orthogonal constraint $\Gbf_k\trans\Gbf_k=\Ibf_{u_k}$. This optimization is over all $r_k\times u_k$ dimensional Grassmann manifolds since $f^{(t)}_k(\Gbf_k)=f^{(t)}_k(\Gbf_k\Obf)$ for any orthogonal matrix $\Obf\in\real{u_k\times u_k}$.

The third step of Algorithm~\ref{algo:all} is to update $\Thetabf$ and $\Omegabf_k$ and $\Omegabf_{0k}$ given the current estimate of the envelope basis $\{ \Gammabf_k \}_{k=1}^m$. We first note that $\Thetabf$ can be estimated by regressing the core tensor, $\Zbf = \llbracket \Ybf; \Gammabf_1\trans, \ldots, \Gammabf\trans_m \rrbracket \in \real{u_1\times \ldots \times u_m}$, on the predictor $\Xbf$ through ordinary least squares without any constraint. That is,
\begin{eqnarray*}
\Thetabf^{(t+1)} = \mathbb{Z}^{(t)} \times_{(m+1)} \left\{ (\mathbb{X} \mathbb{X}\trans)^{-1} \mathbb{X} \right\},
\end{eqnarray*}
where $\Zbf_{i}^{(t)} = \llbracket \Ybf_{i}; (\Gammabf_{1}^{(t+1)})\trans, \ldots, (\Gammabf_{m}^{(t+1)})\trans \rrbracket \in\real{u_{1} \times \ldots \times u_{m}}$, and $\mathbb{Z}^{(t)} \in \real{u_{1} \times \ldots \times u_{m} \times n}$ is the array stacking $\Zbf_1^{(t)}$ to $\Zbf_n^{(t)}$. It is noteworthy that the dimension of the tensor response in this step is reduced from $\prod_{k=1}^{m} r_{k}$ of $\Ybf$ to $\prod_{k=1}^{m} u_{k}$ of $\Zbf$. Next we estimate $\Omegabf_k$, again using the iterative approach of \citet{Dutilleul1999} and \citet{ManceurDutilleul2013}.
\begin{eqnarray*}
\Omegabf_k^{(t+1)} & = & \frac{1}{n\prod_{j\neq k}r_{j}} \sum_{i=1}^{n} \sbf_{i(k)}^{(t)} \left\{ (\Omegabf_{m}^{(t+1)})^{-1} \otimes \ldots\otimes (\Omegabf_{k+1}^{(t+1)})^{-1} \right. \\
 & & \mbox{ } \hspace{1.6in} \left. \otimes (\Omegabf_{k-1}^{(t+1)})^{-1} \otimes \ldots \otimes (\Omegabf_{1}^{(t+1)})^{-1} \right\} \sbf_{i(k)}\trans,
\end{eqnarray*}
where $\sbf_{i}^{(t)} = \Zbf_i^{(t)} - \Thetabf^{(t+1)}\times_{(m+1)}\Xbf_i$ is the residual from the regression of $\Zbf^{(t)}$ on $\Xbf$, and $\sbf_{i(k)}^{(t)}$ is its mode-$k$ matricization. We remark that the above estimation of $\Omegabf_k$ is parallel to the iterative updating of $\Sigmabf_{k}^{(0)}$ during the initialization. The only changes are to replace $\ebf_i$ with $\sbf_i^{(t)}$, to replace $\Sigmabf_{k}^{(0)}$ with $\Omegabf_{k}^{(t+1)}$, and to replace the starting of iteration $\Ibf_{r_k}$ with $\Omegabf_{k}^{(t)}$. Next we estimate $\Omegabf_{0k}$ using the formula,
\begin{eqnarray*}
\Omegabf_{0k}^{(t+1)} & = & \frac{1}{n\prod_{j\neq k}r_{j}} \sum_{i=1}^{n} (\Gammabf_{0k}^{(t+1)})\trans \Ybf_{i(k)} \left\{ (\Sigmabf_{m}^{(t)})^{-1} \otimes \ldots \otimes (\Sigmabf_{k+1}^{(t)})^{-1} \right. \\
 & & \mbox{ } \hspace{1.6in} \left. \otimes (\Sigmabf_{k-1}^{(t)})^{-1} \otimes \ldots \otimes (\Sigmabf_{1}^{(t)})^{-1}\right\} \Ybf_{i(k)}\trans \Gammabf_{0k}^{(t+1)},
\end{eqnarray*}
where $\Gammabf_{0k}^{(t+1)}\in\real{r_k\times (r_k-u_k)}$ is the orthogonal completion of $\Gammabf_{k}^{(t+1)}\in\real{r_k\times u_k}$ such that $(\Gammabf_{k}^{(t+1)},\Gammabf_{0k}^{(t+1)})$ is an orthogonal basis of $\real{r_k}$. We also note that, unlike $\{\Omegabf_{k}\}_{k=1}^{m}$, the estimation of $\{\Omegabf_{0k}\}_{k=1}^{m}$ requires no iteration, since it is only based on the current estimator $\Sigmabf^{(t)}$ and $\Gammabf_k^{(t+1)}$.

Finally, we update $\Bbf$ through its parameterization $\Bbf = \llbracket \Thetabf; \Gammabf_1, \ldots, \Gammabf_m, \Ibf_p \rrbracket$ in Proposition~\ref{thm:parameter}. We are able to obtain the explicit formulae for such an update, i.e.,
\begin{eqnarray*}
\Bbf^{(t+1)} = \mathbb{Y} \times_{1} \PGammak{1}^{(t+1)} \times_{2} \ldots \times_{m} \PGammak{m}^{(t+1)}
 \times_{(m+1)} \left\{ (\mathbb{X} \mathbb{X}\trans)^{-1} \mathbb{X} \right\} =
 \llbracket \Bhatols; \PGammak{1}^{(t+1)}, \ldots, \PGammak{m}^{(t+1)}, \Ibf_p \rrbracket,
\end{eqnarray*}
where $\mathbb{X}$ and $\mathbb{Y}$ are defined in $\Bhatols$ in \eqref{eqn:tensorOLS}. Then we update the covariance $\Sigmabf$ as
\begin{eqnarray*}
\Sigmabf_k^{(t+1)} =  \Gammabf_k^{(t+1)} \Omegabf_k^{(t+1)} (\Gammabf_k^{(t+1)})\trans + \Gammabf_{0k}^{(t+1)} \Omegabf_{0k}^{(t+1)} (\Gammabf_{0k}^{(t+1)})\trans, \textrm{ and }
\Sigmabf^{(t+1)} = \Sigmabf_{m}^{(t+1)} \otimes \ldots \otimes \Sigmabf_{1}^{(t+1)}.
\end{eqnarray*}

\subsection{One-step estimator}
\label{sec:onestep}

Algorithm~\ref{algo:all} is iterative, and steps 2 to 4 are repeated until the objective function converges. Although our numerical experiences suggest that the estimate often does not vary significantly after only a few iterations, the computations involved can still  be intensive. In this iterative procedure, we recognize that the major computational expense arises from Step 2 that estimates the envelope basis $\bigl\{ \Gammabf_k^{(t+1)} \bigr\}_{k=1}^m$ by optimizing the objective functions $f_{k}^{(t)}(\Gbf_k)$ in \eqref{eqn:lik_obj_Gk} over the Grassmann manifolds. This is partly because $f_k^{(t)}$, for $k=1,\dots,m$, depend on each others' minimizers, so that the step of estimation of the envelope basis requires iterations within itself. Moreover, the Grassmann optimization is non-convex and possesses multiple local minima, and as such the algorithm usually requires multiple starting values.

\begin{algorithm}[t]
\begin{algorithmic}
\FOR{$s=0, \ldots, u_k -1$}
\STATE Set $\Gbf_k^{s} = \zerobf$ if $s=0$ and $\Gbf_k^{s} = (\gbf_{k1}, \ldots, \gbf_{ks})$ otherwise
\STATE Construct $\Gbf_{0k}^{s}$ as an orthogonal basis complement to $\Gbf_k^{s}$ in $\real{r_k}$
\STATE Solve the objective function over $\wbf \in \real{r-s}$ subject to $\wbf\trans \wbf = 1$:
\begin{equation*}
\wbf_{k+1} = \arg\min_{\wbf} \log \left\{ \wbf\trans \left( (\Gbf_{0k}^{s})\trans \Sigmabf_k^{(0)} \Gbf_{0k}^{s} \right) \wbf \right\} + \log \left\{ \wbf\trans \left( (\Gbf_{0k}^{s})\trans \Nbf_{k}^{(0)} \Gbf_{0k}^{s} \right)^{-1} \wbf \right \}.
\end{equation*}
\STATE Set $\gbf_{k+1} = \Gbf_{0k}^{s} \wbf_{k+1} \in \real{r_k}$ and normalize to unit length
\ENDFOR
\end{algorithmic}
\caption{\label{algo:1d} Moment-based algorithm for minimizing $f_k^{\ast(0)}(\Gbf_k)$.}
\end{algorithm}

Next we propose an alternative estimator that adopts the same framework of Algorithm~\ref{algo:all}, but replaces Step 2 with an approximate solution by introducing a modified objective function than \eqref{eqn:lik_obj_Gk}. We then restrain the new estimator to be non-iterative, in that it goes through the steps of Algorithm~\ref{algo:all} \emph{only once}. Specifically, the new objective function is of the form,
\begin{equation}\label{eqn:lik_obj_Gk_onestep}
f_k^{\ast(t)}(\Gbf_k) = \log\vert\Gbf\trans_k \Sigmabf_{k}^{(t)} \Gbf_k\vert + \log\vert\Gbf\trans_k (\Nbf_{k}^{(t)})^{-1}\Gbf_k\vert.
\end{equation}
Comparing the two objective functions, the term $\Mbf_{k}^{(t)}$ in \eqref{eqn:lik_obj_Gk} is replaced by $\Sigmabf_{k}^{(t)}$ in \eqref{eqn:lik_obj_Gk_onestep}. It is also interesting to note that, for the first round of iteration, the term $\PGammak{j}^{(t)}$ would take the initial value $\Ibf_{r_j}$, and as such the term $\deltabf_i^{(t)}$ becomes the OLS residual $\ebf_i = \Ybf_i - \Bbf^{(0)} \times_{(m+1)} \Xbf_i$. Consequently, $\Mbf_k^{(0)}=\Sigmabf_k^{(0)}$, and thus $f_k^{\ast(0)}(\Gbf_k)=f_k^{(0)}(\Gbf_k)$. This new objective function \eqref{eqn:lik_obj_Gk_onestep} has some appealing features. First of all, estimation of the envelope basis $\Gammabf_{k}$ through $f_k^{\ast(t)}$ does not depend on the values of other envelope basis $\Gammabf_{j}$, $j\neq k$. Therefore, estimation of $\Gammabf_1$ to $\Gammabf_m$ becomes \emph{separable}. This property alone could increase computational efficiency substantially. Second, the optimization of $f_k^{\ast(t)}$ can be achieved by the sequential algorithm recently proposed by \citet{CookZhang2014algorithm}, which is much faster and more stable than the Grassmann optimization, and the estimation result does not hinge on the initial guess. For completeness of the presentation, we summarize this sequential algorithm in Algorithm~\ref{algo:1d}.

Thanks to both the modified objective function as well as the non-iterative fashion of the optimization, the resulting one-step estimator is computationally much faster than the iterative estimator. Our numerical studies have found that it has a competitive finite sample performance. Moreover, as we show in the next section, like its iterative counterpart, this one-step estimator remains a consistent estimator of the true parameters. Therefore, we recommend the one-step estimator in practice.

\section{Asymptotics}
\label{sec:theory}

In this section we study the asymptotic properties of the envelope based estimators as the sample size goes to infinity. We investigate both the iterative estimator, denoted as $\Bhatenvit$, and the one-step estimator, denoted as $\Bhatenvos$, under two scenarios: the normality of the error distribution holds or does not hold.

\subsection{Consistency}

We first establish that, under fairly weak moment conditions, both the iterative estimator and the one-step estimator are $\sqrt{n}$-consistent, when the error term in the tensor response linear model \eqref{eqn:model-tensor} does not necessarily follow a  normal distribution. We note that the consistency is established in terms of the projection matrices, $\PGammak{k}^{it}$ for the iterative estimator $\Bhatenvit$, and $\PGammak{k}^{os}$ for the one-step $\Bhatenvos$, since a subspace can have infinitely many semi-orthogonal basis but only one unique projection matrix.

\begin{theorem} \label{thm:gamma-consistency}
Assuming $\vect(\varepsilonbf_i)$, $i=1,\dots,n$, in model \eqref{eqn:model-tensor} are i.i.d.\ with finite fourth moments, then the projection $\PGammak{k}^{it}$ and $\PGammak{k}^{os}$ are both $\sqrt{n}$-consistent estimators for the projection onto the envelope $\envsk{k}$. The corresponding estimators $\Bhatenvit$ and $\Bhatenvos$ both converge at rate-$\sqrt{n}$ to the true tensor coefficient $\Bbf_{\textrm{{\tiny TRUE}}}$.
\end{theorem}

\subsection{Asymptotic normality}

We next establish the asymptotic normality of the iterative estimator $\Bhatenvit$ when the error term $\vect(\varepsilonbf)$ follows a normal distribution. Since only the iterative estimator is to be examined, we abbreviate its notation simply as $\Bhatenv$, and the corresponding projection $\PGammak{k}^{it}$ as $\PGammak{k}$. Under this condition, this envelope based estimator becomes the maximum likelihood estimator (MLE).

\begin{theorem} \label{thm:asympt_eff}
Assuming $\vect(\varepsilonbf_i)$, $i=1,\dots,n$, in model \eqref{eqn:model-tensor} are i.i.d.\ with a normal distribution, then the projection $\PGammak{k}$ is the MLE for the projection onto the envelope $\envsk{k}$. The corresponding estimator $\Bhatenv$ is the MLE, and $\sqrt{n}\vect(\Bhatenv - \Bbf_{\textrm{{\tiny TRUE}}}) \rightarrow N(0,\Ubf_{\textrm{{\tiny ENV}}})$. Moreover, the OLS estimator $\Bhatols$ satisfies that $\sqrt{n}\vect(\Bhatols - \Bbf_{\textrm{{\tiny TRUE}}}) \rightarrow N(0,\Ubf_{\textrm{{\tiny OLS}}})$, and $\Ubf_{\textrm{{\tiny ENV}}} \leq \Ubf_{\textrm{{\tiny OLS}}}$.
\end{theorem}

In addition to the established asymptotic normality, an important finding of Theorem~\ref{thm:asympt_eff} is that the asymptotic variance of the envelope estimator $\Bhatenv$ is no greater than that of the OLS estimator $\Bhatols$. Therefore, $\Bhatenv$ is asymptotically more efficient than $\Bhatols$. One can conveniently obtain the asymptotic covariance of $\vect(\Bhatols)$,
\begin{eqnarray*}
\Ubf_{\textrm{{\tiny OLS}}} = \Sigmabf_{\mathbf{X}}^{-1}\otimes\Sigmabf = \Sigmabf_{\mathbf{X}}^{-1}\otimes\Sigmabf_m\otimes\cdots\otimes\Sigmabf_1,
\end{eqnarray*}
where $\Sigmabf_{\mathbf{X}} = \cov(\Xbf)$. Next we give two additional results to gain more insight of $\Ubf_{\textrm{{\tiny ENV}}}$. One assumes that the envelope basis is known a priori, and the other obtains the asymptotic variance of the estimator for both $\Bbf$ and $\Sigmabf$ jointly.

We first assume the envelope basis is known, and denote the corresponding envelope estimator of $\Bbf$ as $\widehat \Bbf_{\mathbf{\Gamma}}$. We then compare its asymptotic variance with that of $\Bhatols$.

\begin{theorem}
Under the same conditions as in Theorem~\ref{thm:asympt_eff}, $\widehat \Bbf_{\mathbf{\Gamma}}$ is $\sqrt{n}$-consistent and asymptotically normal. The asymptotic covariance of $\vect(\widehat \Bbf_{\mathbf{\Gamma}})$ is
\begin{eqnarray*}
\Ubf_{\mathbf{\Gamma}} = \Sigmabf_{\mathbf{X}}^{-1}\otimes\PGammak{m}\Sigmabf_m\PGammak{m}\otimes
\cdots\otimes\PGammak{1}\Sigmabf_1\PGammak{1} = \Sigmabf_{\mathbf{X}}^{-1}\otimes\Gammabf_m\Omegabf_m\Gammabf_m\trans\otimes \cdots\otimes\Gammabf_1\Omegabf_1\Gammabf_1\trans.
\end{eqnarray*}
\end{theorem}

\noindent
Recall the decomposition $\Sigmabf_k = \Gammabf_k \Omegabf_k \Gammabf_k\trans + \Gammabf_{0k} \Omegabf_{0k} \Gammabf_{0k}\trans$, $k = 1, \ldots, m$, in Proposition~\ref{thm:parameter}. Then it is straightforward to see that $\Ubf_{\mathbf{\Gamma}} \leq \Ubf_{\textrm{{\tiny OLS}}}$, and the more dominating the immaterial variation $\Gammabf_{0k} \Omegabf_{0k} \Gammabf_{0k}\trans$ compared to the material variation $\Gammabf_k \Omegabf_k \Gammabf_k\trans$, the bigger the difference is between $\Ubf_{\mathbf{\Gamma}}$ and $\Ubf_{\textrm{{\tiny OLS}}}$. This result agrees with the pattern we have observed and reviewed in Section~\ref{sec:vector-resp} for the vector response case, and shows the explicit gain of the envelope estimator in terms of estimation efficiency.

We next compare the asymptotic covariance of the envelope estimator and the OLS estimator when the envelope basis is unknown. Intuitively, there is an extra term in the covariance of the envelope estimator as the cost of estimating the unknown envelope basis. Toward that end, we introduce the following notations.
\begin{equation*}\label{eqn:model-parameters}
\hbf=
\left(\begin{array}{c}
\hbf_1\\
\hbf_2
\end{array}\right)
=
\left(\begin{array}{c}
\vect(\Bbf)\\
\vech(\Sigmabf)
\end{array}\right),
\quad
\phibf=
\left(\begin{array}{c}
\phibf_1\\
\phibf_2\\
\vdots\\
\phibf_{m+1}\end{array}
\right)
=
\left(\begin{array}{c}
\vect(\Bbf)\\
\vech(\Sigmabf_1)\\
\vdots\\
\vech(\Sigmabf_{m})\end{array}
\right)
,\quad
\xibf
=
\left(
\begin{array}{c}
\xibf_1\\
\vdots\\
\xibf_{3m+1}
\end{array}
\right),
\end{equation*}
where the operator $\vech(\cdot):\real{r\times r}\mapsto\real{r(r+1)/2}$ stacks the unique entries of a symmetric matrix into a column vector, $\xibf_1=\vect(\Thetabf)$, $\{\xibf_j\}_{j=2}^{m+1}=\{\vect(\Gammabf_k)\}_{k=1}^{m}$, $\{\xibf_j\}_{j=m+2}^{2m+1}=\{\vech(\Omegabf_k)\}_{k=1}^{m}$, and $\{\xibf_j\}_{j=2m+2}^{3m+1}=\{\vech(\Omegabf_{0k})\}_{k=1}^{m}$. It is interesting to note that the total number of free parameters is reduced from $\hbf$ to $\phibf$ by $\prod_{k=1}^{m} r_k (\prod_{k=1}^{m} r_k+1)/2-\sum_{k=1}^{m}r_k(r_k+1)/2$ because of the imposed separable Kronecker covariance structure, and is further reduced from $\phibf$ to $\xibf$ by $p(\prod_{k=1}^{m}r_k - \prod_{k=1}^{m} u_k)$. We also note that $\hbf$ is an estimable functions of $\phibf$ and $\xibf$, respectively, and thus we write $\hbf = \hbf(\phibf) = \hbf(\xibf)$. Let $\Jbf_{\hbf}$ denote the Fisher information matrix for $\hbf$, let $\Hbf = {\partial \hbf(\phibf)}/{\partial \phibf}$, and $\Kbf = {\partial \hbf(\xibf)}/{\partial \xibf}$. The explicit forms of $\Jbf_{\hbf}, \Hbf$ and $\Kbf$ are given in the Supplementary Materials. Furthermore, let $\hbf_{\textrm{{\tiny OLS}}}$ denote the OLS estimator of $\hbf$, $\hbf_{\textrm{{\tiny ENV}}}$ be the envelope estimator, and $\hbf_{\textrm{{\tiny TRUE}}}$ be the true parameter. Then, we have the following result.

\begin{theorem}\label{thm:asymptotics_overall}
Under the same conditions as in Theorem~\ref{thm:asympt_eff}, both $\hbf_{\textrm{{\tiny OLS}}}$ and $\hbf_{\textrm{{\tiny ENV}}}$ are $\sqrt{n}$-consistent and asymptotically normal, so that $\sqrt{n}(\hbf_{\textrm{{\tiny OLS}}} - \hbf_{\textrm{{\tiny TRUE}}}) \rightarrow N(0, \Vbf_{\textrm{{\tiny OLS}}})$, where $\Vbf_{\textrm{{\tiny OLS}}} = \Hbf(\Hbf\trans\Jbf_{\hbf}\Hbf)^{\dagger}\Hbf\trans$, and $\sqrt{n}(\hbf_{\textrm{{\tiny ENV}}} - \hbf_{\textrm{{\tiny TRUE}}}) \rightarrow N(0, \Vbf_{\textrm{{\tiny ENV}}})$, where $\Vbf_{\textrm{{\tiny ENV}}} = \Kbf(\Kbf\trans\Jbf_{\hbf}\Kbf)^{\dagger}\Kbf\trans$.
Moreover,
\begin{eqnarray*}
\Vbf_{\textrm{{\tiny OLS}}} - \Vbf_{\textrm{{\tiny ENV}}}
=  \Jbf_{\hbf}^{-1/2}\left( \Pbf_{ \Jbf_{\hbf}^{1/2}\Hbf}- \Pbf_{ \Jbf_{\hbf}^{1/2}\Kbf}\right) \Jbf_{\hbf}^{-1/2}
=  \Jbf_{\hbf}^{-1/2} \Pbf_{\Jbf_{\hbf}^{1/2}\Hbf} \Qbf_{\Jbf_{\hbf}^{1/2} \Kbf} \Jbf_{\hbf}^{-1/2} \geq 0.
\end{eqnarray*}
\end{theorem}

\noindent
Once again, the envelope estimator is asymptotically more efficient than the OLS estimator. On the other hand, the envelope estimator of $\Bbf$ and $\Sigmabf$ are asymptotically \emph{correlated}. As such, there is no explicit form for the asymptotic covariance of $\Bhatenv$, except that it is the $p(\prod r_k) \times p(\prod r_k)$ upper-left block of $\Vbf_{\textrm{{\tiny ENV}}}$, when the envelope basis is unknown. This is different from the vector response case.

In applications such as brain imaging analysis, it is often useful to produce a voxel-by-voxel $p$-value map, so one can visually identify subregions of brains that display distinctive patterns between disease and control groups. Given the results of Theorems~\ref{thm:asympt_eff} and \ref{thm:asymptotics_overall}, we can produce such a $p$-value map for our envelope based estimator $\Bhatenv$. In principle, one can obtain its asymptotic covariance $\Ubf_{\textrm{{\tiny ENV}}}$ by extracting the upper-left block of $\Vbf_{\textrm{{\tiny ENV}}}$. In practice, however, we suggest to substitute $\Ubf_{\textrm{{\tiny ENV}}}$ with $\Ubf_{\textrm{{\tiny OLS}}}$, which is computationally much simpler, though it would lead to more conservative $p$-values. Once the $p$-values are obtained, one can further employ either simple thresholding or false discovery rate correction.

\section{Simulations}
\label{sec:simulation}

In this section, we report simulations to compare the new estimator with two major competitors, the one-at-a-time OLS estimator (Section~\ref{sec:comp-ols}), and the tensor predictor regression of \citet{ZhouLiZhu2013} (Section~\ref{sec:comp-CP}). It is noteworthy that, in the first comparison, the data was simulated from a model that conforms with the envelope structure, whereas in the second comparison, the model does \emph{not} follow this structure. Therefore, the numerical experiment also demonstrates the performance of our estimator under model misspecification. Moreover, we investigate the effect of the envelope dimension and of the magnitude of immaterial variation on the coefficient estimation (Section~\ref{sec:dim}), and examine the case when the response is a three-way tensor (Section~\ref{sec:order3}).

\subsection{Comparison with OLS estimator}
\label{sec:comp-ols}

\begin{figure}[t]
\includegraphics[height=4.1in]{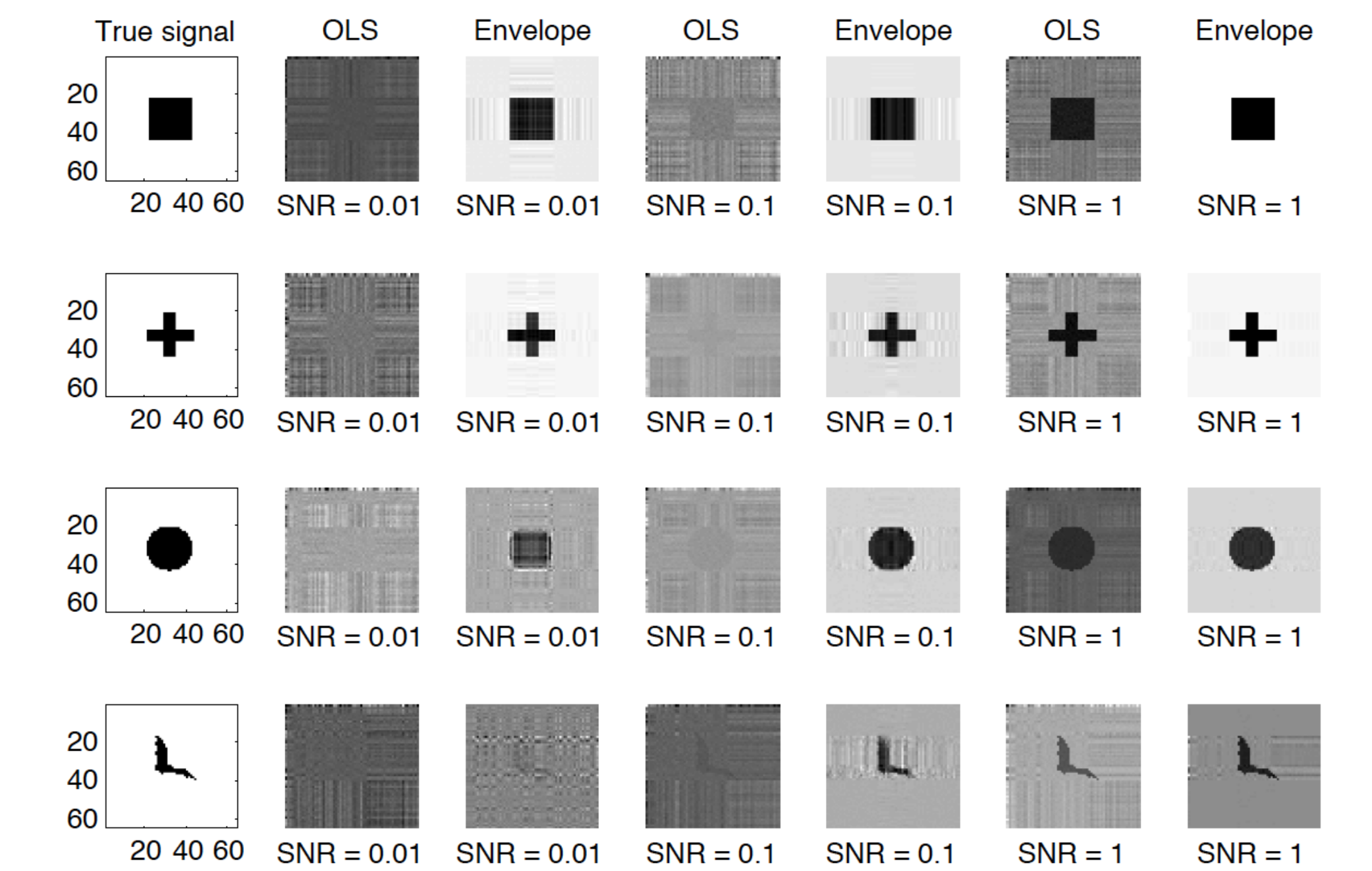}
\caption{Comparison with OLS: The true and estimated regression coefficient tensors under various signal shapes and signal-to-noise ratios (SNR). \label{fig:compols}}
\end{figure}

We begin with a comparison with the alternative solution that dominates the literature, the OLS estimator, which fits one response component at a time. Specifically, we consider the model of the form \eqref{eqn:model-tensor},
\begin{eqnarray} \label{eqn:model-tensor-sim}
\Ybf_{i} = \Bbf X_{i} + \sigma \varepsilonbf_{i}, \;\;\; i=1,\dots,n.
\end{eqnarray}
We set the sample size $n=20$, a fairly small value, to mimick the common scenario of imaging studies with a small number of subjects. In this model, $\Ybf_i$ is a $64 \times 64$ matrix, $X_i$ is a scalar that only takes two values, 0 or 1, representing for instance the disease and control groups. $\Bbf \in \real{64 \times 64}$ represents the mean change of the response between the two groups. Elements of $\Bbf$ are either 0 or 1, and is plotted in the first column of Figure~\ref{fig:compols}. We varied the shape of $\Bbf$ among a square, a cross, a round disk and a bat shape. The constant $\sigma$ in front of the error term $\varepsilonbf$ was introduced to explicitly control the signal strength, and it took a value such that the signal-to-noise ratio (SNR) equals 0.01, 0.1, and 1, respectively. The error $\varepsilonbf$ was generated from a matrix normal distribution, $\vect(\varepsilonbf)\sim N(0,\Sigmabf_2 \otimes \Sigmabf_1)$. To make the model conform to the generalized sparsity principle \eqref{eqn:assumption-tensor}, we generated $\Sigmabf_k = \Gammabf_{k} \Omegabf_{k}\Gammabf_{k}\trans + \Gammabf_{0k} \Omegabf_{0k} \Gammabf_{0k}\trans$, $k = 1, 2$, then normalized it by its Frobenius norm. We set the working envelope dimension $u_1 = u_2$ equal to the numerical dimension of the true coefficient matrix $\Bbf$, which is 1 for the square signal, 2 for the cross, 8 for the disk, and 14 for the bat-shape. We eigen-decomposed the coefficient matrix as $\Bbf = \Gbf \Dbf \Gbf\trans$. Then we generated $\Gammabf_k = \Gbf \Obf_k$, for an orthogonal matrix $\Obf_k \in \real{u_k \times u_k}$, whose elements were from a standard uniform distribution. This way, it is guaranteed that $\spn(\Bbf) \subseteq \spn(\Gammabf_1)$ and $\spn(\Bbf\trans) \subseteq \spn(\Gammabf_2)$. We then orthogonalized $\Gammabf_k$ and obtained $\Gammabf_{0k}$. The covariances $\Omegabf_k \in \real{u_k \times u_k}$ and $\Omegabf_{0k} \in \real{(r_k-u_k) \times (r_k-u_k)}$ were generated of the form $\Abf \Abf\trans$, where $\Abf$ is a square matrix with matching dimension and all its elements were from a standard uniform distribution.

Figure~\ref{fig:compols} summarizes the estimated coefficient matrix $\Bbf$, under varying image shapes and signal strengths. It is clearly seen that the envelope estimator $\Bhatenv$ substantially outperforms the one-at-a-time OLS estimator $\Bhatols$. For instance, when the signal is weak (SNR is 0.01 or 0.1), the OLS estimator failed to identify any meaningful signal, whereas the envelope estimator produced a sound recovery. Moreover, we emphasize that such a performance is achieved under a very small sample size ($n=20$).

\subsection{Comparison with tensor predictor regression}
\label{sec:comp-CP}

Next we compare our method with a recent proposal of tensor regression \citep{ZhouLiZhu2013} that studies association between a scalar response and a \emph{tensor predictor}. Even though both methods are motivated from neuroimaging analysis, and both involve a tensor variable in a regression analysis, the two clearly differ in the role the tensor plays in regression and the corresponding interpretation. Moreover, the techniques involved differ significantly too, with our approach utilizing the generalized sparsity principle, whereas \citet{ZhouLiZhu2013} employed a reduced-rank decomposition, the canonical decomposition or parallel factors (CP decomposition), of the coefficient tensor.

We consider the model of \citet{ZhouLiZhu2013},
\begin{eqnarray} \label{eqn:model-CP-sim}
Y_i = \langle \Bbf,\Xbf_i \rangle + \varepsilon_i, \;\;\; i=1,\ldots,n,
\end{eqnarray}
where $Y_i$ is a scalar response, $\Xbf \in \real{64 \times 64}$ is an image whose elements were all drawn from a standard normal distribution, and the error $\epsilon$ is standard normal and independent of $\Xbf$. The coefficient matrix $\Bbf \in \real{64 \times 64}$ was set in the same way as in Section~\ref{sec:comp-ols}. $\langle \Bbf, \Xbf_i \rangle = \langle \vect(\Bbf), \vect(\Xbf) \rangle$ is the tensor inner product. We examined three sample sizes, $n = 300, 900$ and $2700$, respectively.

\begin{figure}[t]
\includegraphics[height=3.975in]{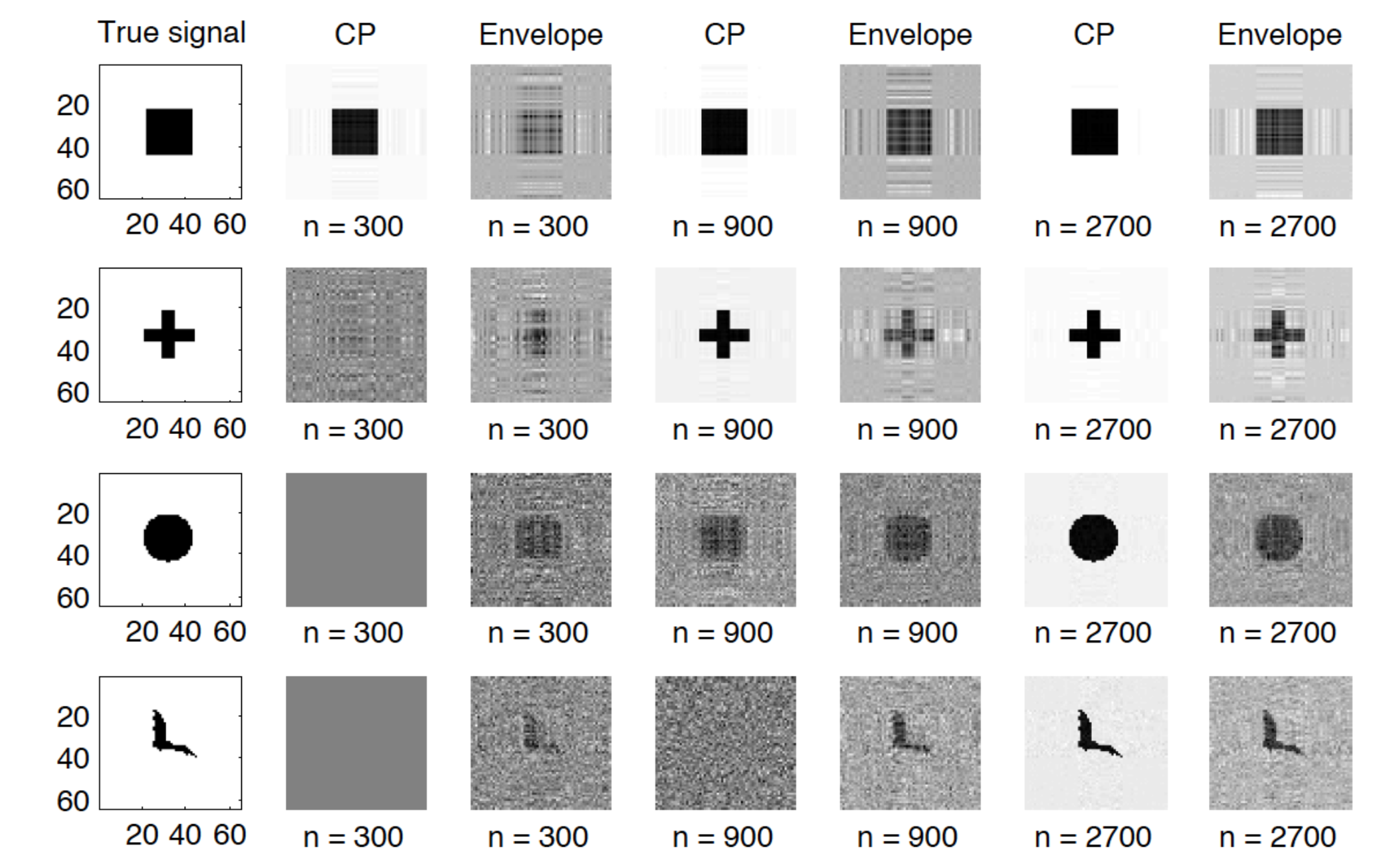}
\caption{Comparison with tensor predictor regression: The true and estimated regression coefficient tensors under various signal shapes and sample sizes. \label{fig:compcp}}
\end{figure}

Figure~\ref{fig:compcp} summarizes the results. It is interesting to observe that the enveloped based estimator outperforms the CP estimator of \citet{ZhouLiZhu2013} when the sample size is small ($n=300$) to moderate ($n=900$), and underperforms only when the sample size is fairly large ($n=2700$) but still produces a reasonable signal recovery. It is also noteworthy that, in this example, the data was generated from model \eqref{eqn:model-CP-sim}, based upon which that the CP estimator was built. As a result, it actually favors the CP estimator. On the other hand, the generalized sparsity principle \eqref{eqn:assumption-tensor} is not guaranteed in this setup. Therefore this comparison shows the promise of our envelope estimator even when the assumed envelope structure does not hold in the data.

\subsection{Envelope dimension and immaterial information}
\label{sec:dim}

We next investigate two issues: the effect of the working envelope dimension, $\{u_k\}_{k=1}^{m}$, which are the tuning parameters of our method, and the effect of magnitude of the immaterial information on our proposed envelope based estimation.

\begin{figure}[t]
\begin{center}
\begin{tabular}{c}
\vspace{-0.15in}
(a) Immaterial-to-material variation ratio $\sigma_0^2=1$ \\
\includegraphics[bb=100bp 300bp 512bp 500bp, scale=0.925]{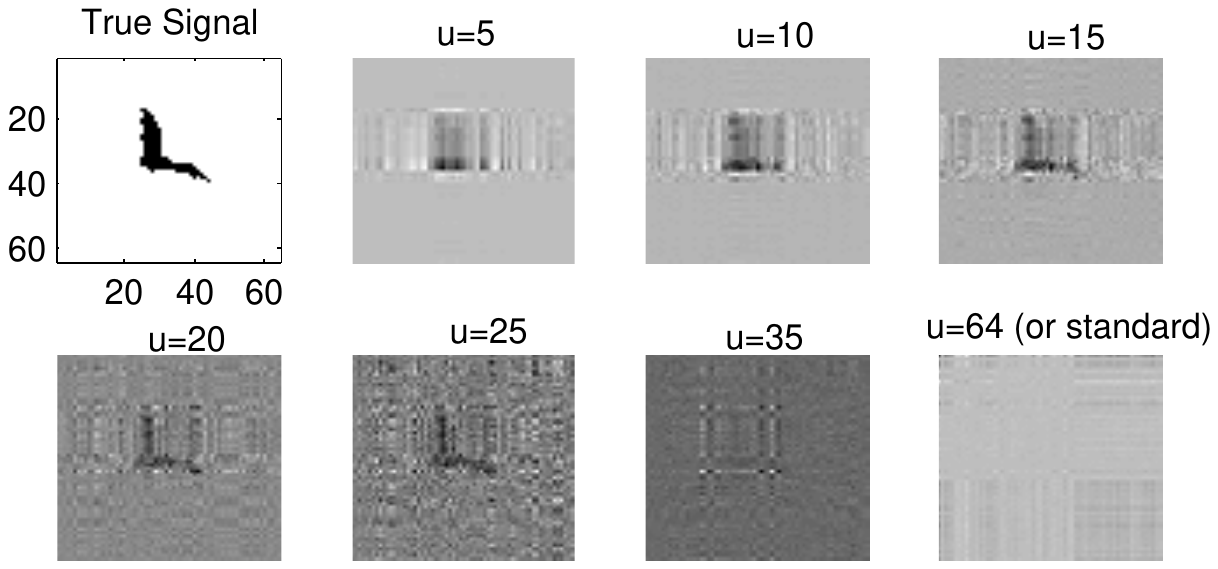} \\
\vspace{-0.15in}
(b) Immaterial-to-material variation ratio $\sigma_0^2=1000$ \\
\includegraphics[bb=100bp 300bp 512bp 500bp, scale=0.925]{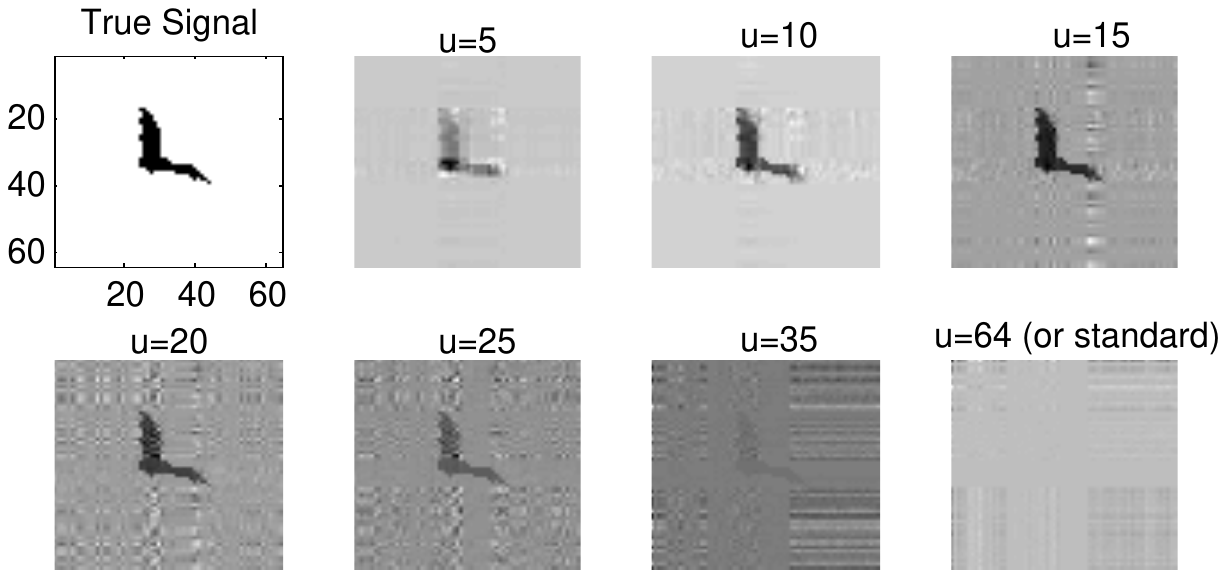}
\end{tabular}
\end{center}
\vspace{-0.3in}
\caption{Effect of working envelope dimension and the strength of immaterial information on the envelope based estimator.
\label{fig:dim}}
\end{figure}

For the first problem, we generated $n=100$ i.i.d.\ samples from model \eqref{eqn:model-tensor-sim} with the bat-shape signal, which is a natural shape and is relatively more complicated than the geometric shapes. We then varied the working envelope dimension $u_1 = u_2 = u$, with $u = \{ 5, 10, 15, 20, 25, 35, 64\}$, while the numerical rank of the bat-shape signal equals 14 in this example. We also note that, if one sets the working envelope dimension $u_k$ the same as the dimension of the tensor response $r_k$, then the envelope estimator degenerates to the OLS estimator. Figure~\ref{fig:dim}(a) shows one snapshot of the results. We first see that, all the envelope estimators ($u < 64$) outperformed the OLS estimator ($u = 64$), reinforcing the pattern observed in Section~\ref{sec:comp-ols}. When the working envelope dimension is smaller than the true signal dimension ($u < 14$ in this example), the envelope estimator produced reasonable but mediocre recovery. When the working dimension exceeds the truth ($u > 14$), the envelope estimator produced much refined recovery. Meanwhile, as the working dimension increases, there is a sign of overfitting, but the quality of the recovered signal remains competitive.

For the second problem, we continued to employ model \eqref{eqn:model-tensor-sim}, but introduced an additional scalar $\sigma_0$ in the covariance $\Sigmabf_{k} = \Gammabf_{k} \Omegabf_{k} \Gammabf_{k}\trans + \sigma_{0}^{2} \; \Gammabf_{0k}\Omegabf_{0k} \Gammabf_{0k}\trans$, where $\sigma_0$ controls the magnitude of the immaterial information. Figure~\ref{fig:dim}(b) shows one snapshot of the results when $\sigma_0^2 = {1000}$, whereas Figure~\ref{fig:dim}(a) had $\sigma_0^2 = 1$. Comparing the two figures, we first verify that, the more dominant of the immaterial information (i.e., the larger value of $\sigma_0$), the better performance of the envelope estimator. On the other hand, the OLS estimator continued to fail to identify any meaningful signal. This  demonstrates the importance of recognizing the immaterial information to improve the estimation.

\subsection{Three-way tensor}
\label{sec:order3}

\begin{table}[b]
\begin{center}
\begin{tabular}{|c|c|c|c|c|c|}
\hline
\multicolumn{1}{|c}{} &  & \multicolumn{2}{c|}{$\Vert\Bhatols-\Bbf\Vert^2$} & \multicolumn{2}{c|}{$\Vert\Bhatenv-\Bbf\Vert^{2}$}\tabularnewline
\cline{3-6}
\multicolumn{1}{|c}{} &  & Average & S.E. & Average & S.E.\tabularnewline
\hline
\hline
\multirow{2}{*}{$(u_{1},u_{2},u_{3})=(2,3,4)$} & $n=100$ & 127 & 0.07 & 4.17 & 0.04\tabularnewline
\cline{2-6}
 & $n=400$ & 29.0 & 0.03 & 0.81 & 0.01\tabularnewline
\hline
\multirow{2}{*}{$(u_{1},u_{2},u_{3})=(5,5,5)$} & $n=100$ & 133 & 0.03 & 3.57 & 0.01\tabularnewline
\cline{2-6}
 & $n=400$ & 32.2 & 0.05 & 0.69 & 0.01\tabularnewline
\hline
\multirow{2}{*}{$(u_{1},u_{2},u_{3})=(10,10,10)$} & $n=100$ & 213 & 3.68 & 4.08 & 0.40\tabularnewline
\cline{2-6}
 & $n=400$ & 51.8 & 1.98 & 0.89 & 0.20\tabularnewline
\hline
\end{tabular}
\par
\end{center}

\caption{Average and standard error (in parenthesis) of the estimation accuracy measured by $\Vert \widehat \Bbf - \Bbf\Vert^2$ based on 100 data replications.\label{tab:simulation3d}}
\end{table}

In the above simulations, we have primarily focused on the matrix-valued response (order-2 tensor) and a single predictor, since it enables a direct visualization of the coefficient estimator. In this section, we considered a tensor response model where the response becomes an order-3 tensor with dimensions $(r_{1},r_{2},r_{3}) = (20, 30, 40)$, and the predictor is a $5$-dimensional vector. The rest of the setup was similar to that in Section~\ref{sec:comp-ols}. We simulated data with different envelope dimensions: $(u_{1},u_{2},u_{3})=(2,3,4)$, $(5,5,5)$ or $(10,10,10)$, and different sample sizes: $n=100$ or $400$. For each combination, we simulated 100 data replications, and report the average and the standard error of $\Vert\Bhatols-\Bbf\Vert^2$ and $\Vert\Bhatenv-\Bbf\Vert^2$ in Table~\ref{tab:simulation3d}. Again, the envelope based estimator showed a dramatic improvement over the OLS estimator in terms of estimation accuracy.

\section{Real Data Analysis}
\label{sec:realdata}

\subsection{EEG data analysis}
\label{sec:EEG}

\begin{figure}[t!]
\begin{center}
\includegraphics[bb= 50bp 200bp 512bp 670bp, scale=0.775]{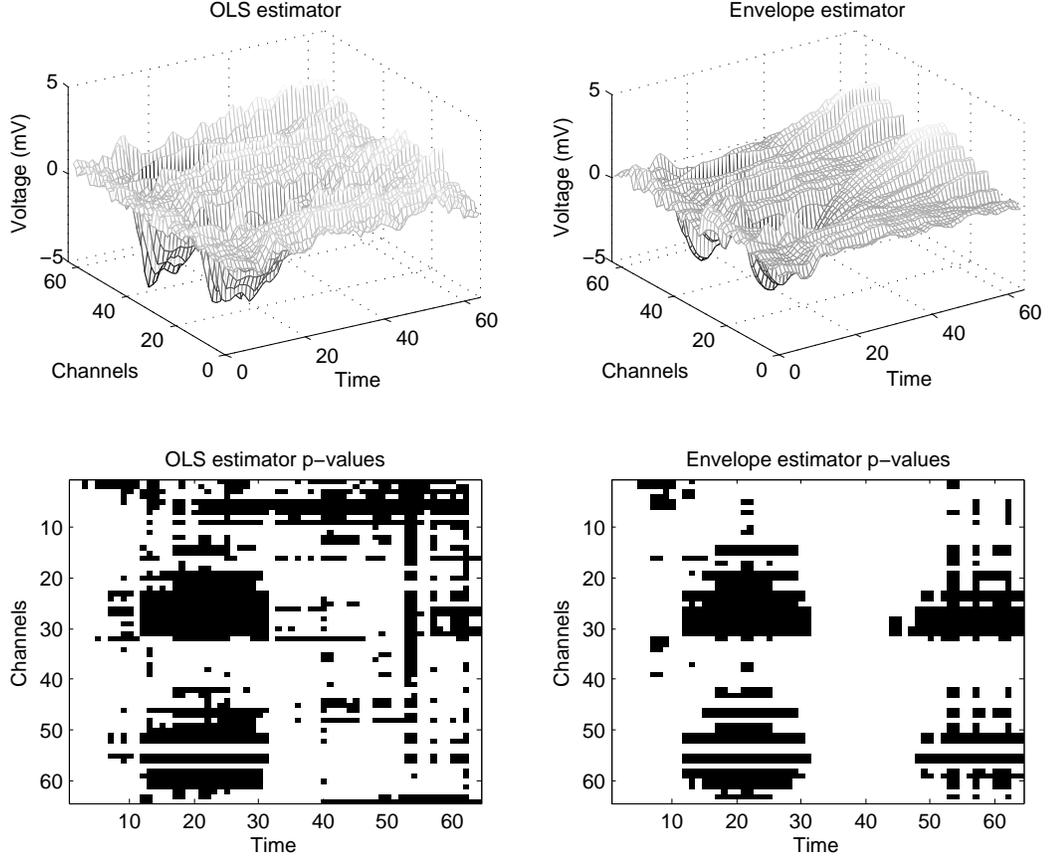}
\end{center}
\vspace{-0.05in}
\caption{EEG data analysis: top panels are estimated regression coefficients; bottom panels are the $p$-value maps, where the black regions correspond to the $p$-value less than the threshold 0.05. \label{fig:eeg}}
\end{figure}

We first analyzed an electroencephalography (EEG) data for an alcoholism study. The data was obtained from \url{https://archive.ics.uci.edu/ml/datasets/EEG+Database}. It contains 77 alcoholic individuals and 44 controls. Each individual was measured with 64 electrodes placed on the scalp sampled at 256 Hz for one second, resulting an EEG image of 64 channels by 256 time points. More information about data collection and some analysis can be found in \citet{Zhang_etal1995} and \citet{li2010dimensionfolding}. To facilitate the analysis, we downsized the data along the time domain by averaging four consecutive time points, yielding a $64 \times 64$ matrix response. We report both the OLS estimator and the envelope based estimator in Figure~\ref{fig:eeg}, where the upper panels show the coefficient estimator and the lower panels show the corresponding $p$-value maps, thresholded at 0.05. It is interesting to observe that, the envelope estimator identifies the channels between about 15 to 30, and between 45 to 60, at time range from 30 to 120, and from about 200 to 240, are mostly relevant to distinguish the alcoholic group from the control. By contrast, the OLS estimator is much more variable, and  the revealed signal regions are much less clear.

\subsection{ADHD data analysis}
\label{sec:adhd}

\begin{figure}[t!]
\begin{center}
\begin{tabular}{ccccc}
%\multicolumn{2}{c}{OLS estimator} &\multicolumn{1}{c}{} & \multicolumn{2}{c}{Envelope estimator} \\
\includegraphics[bb=220bp 0bp 750bp 520bp,scale=0.2]{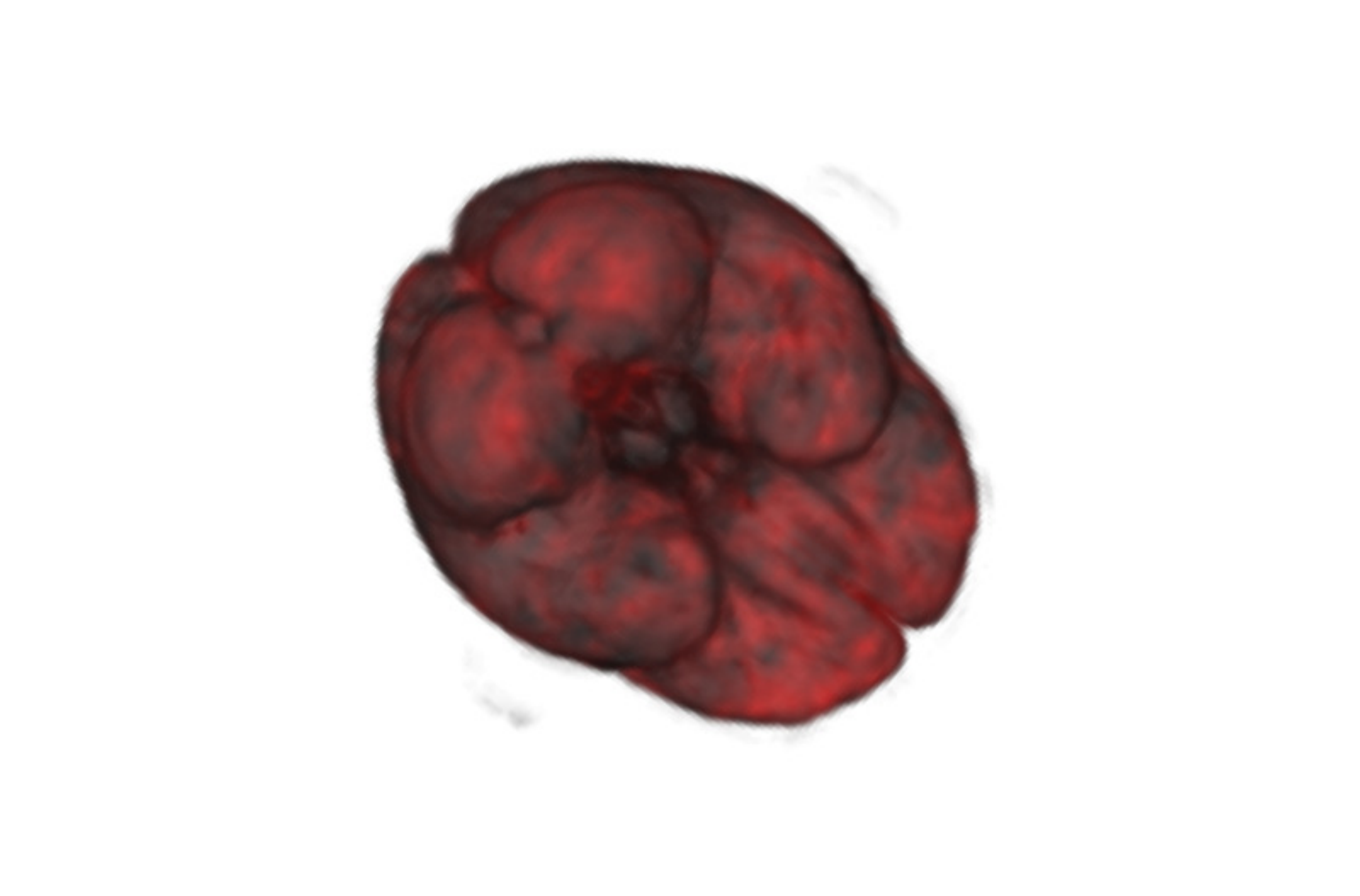}    &
\includegraphics[bb=220bp 0bp 750bp 520bp,scale=0.2]{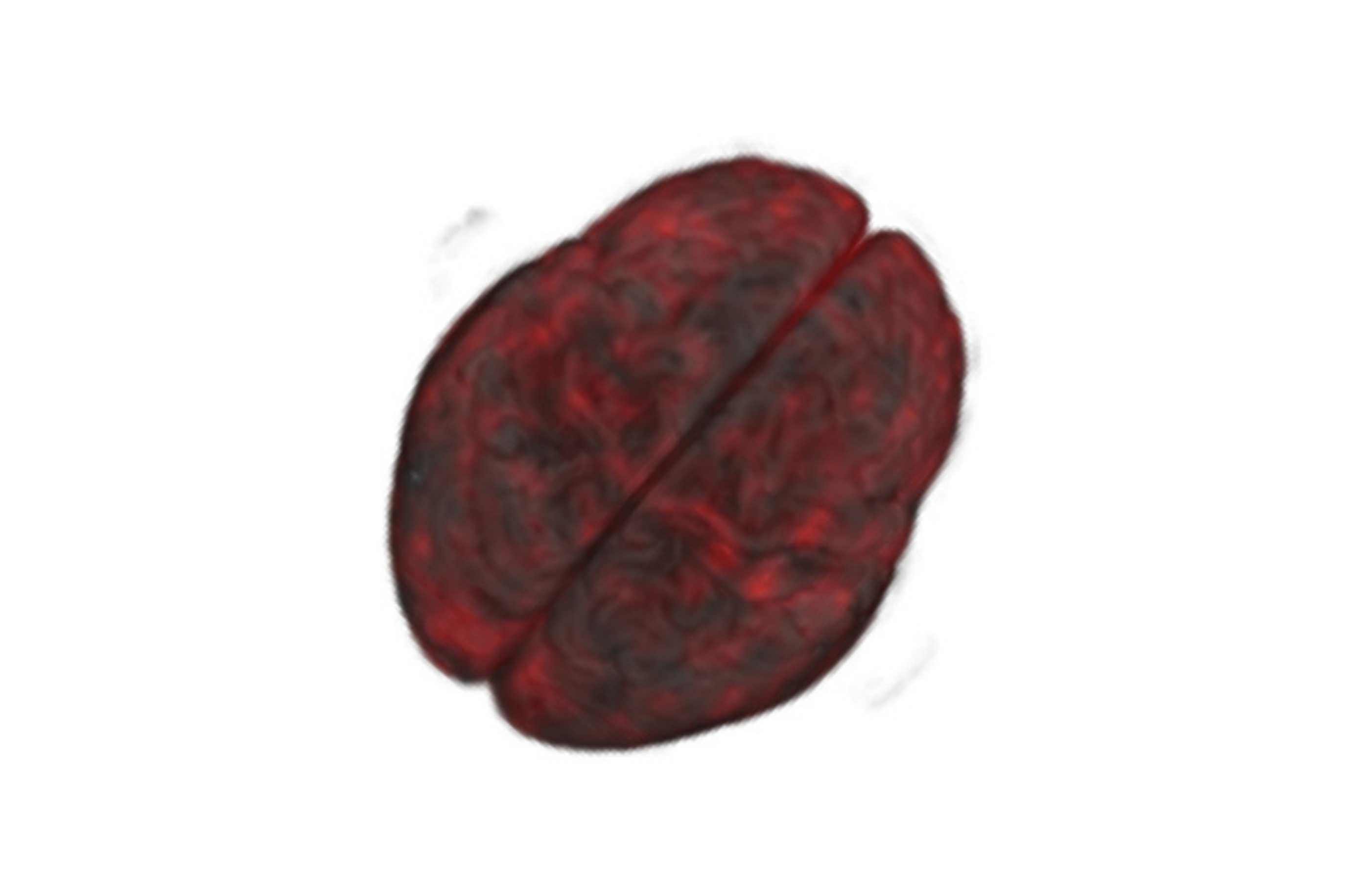}  & \mbox{ } &
\includegraphics[bb=220bp 0bp 750bp 520bp,scale=0.2]{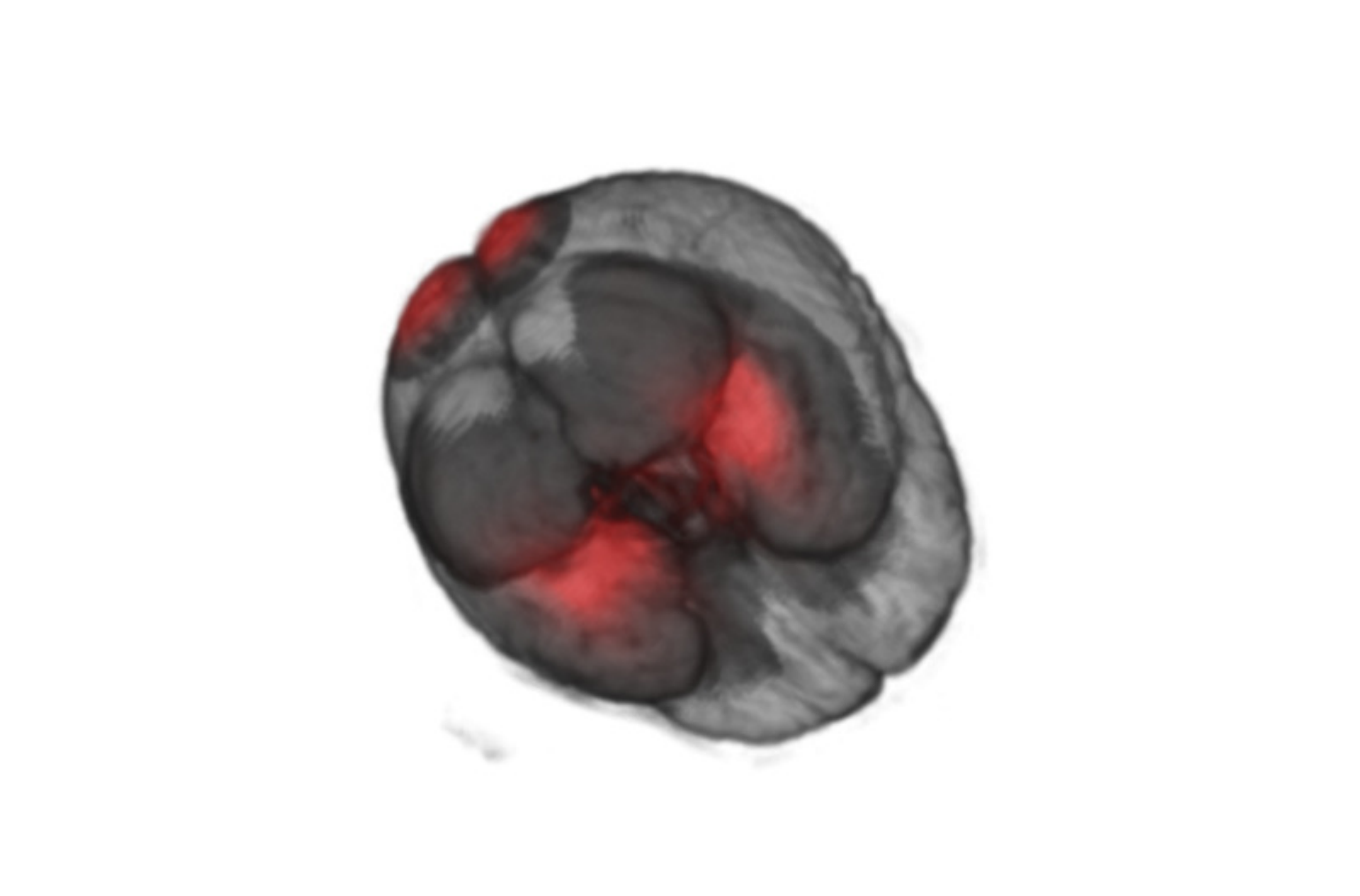}   &
\includegraphics[bb=220bp 0bp 750bp 520bp,scale=0.2]{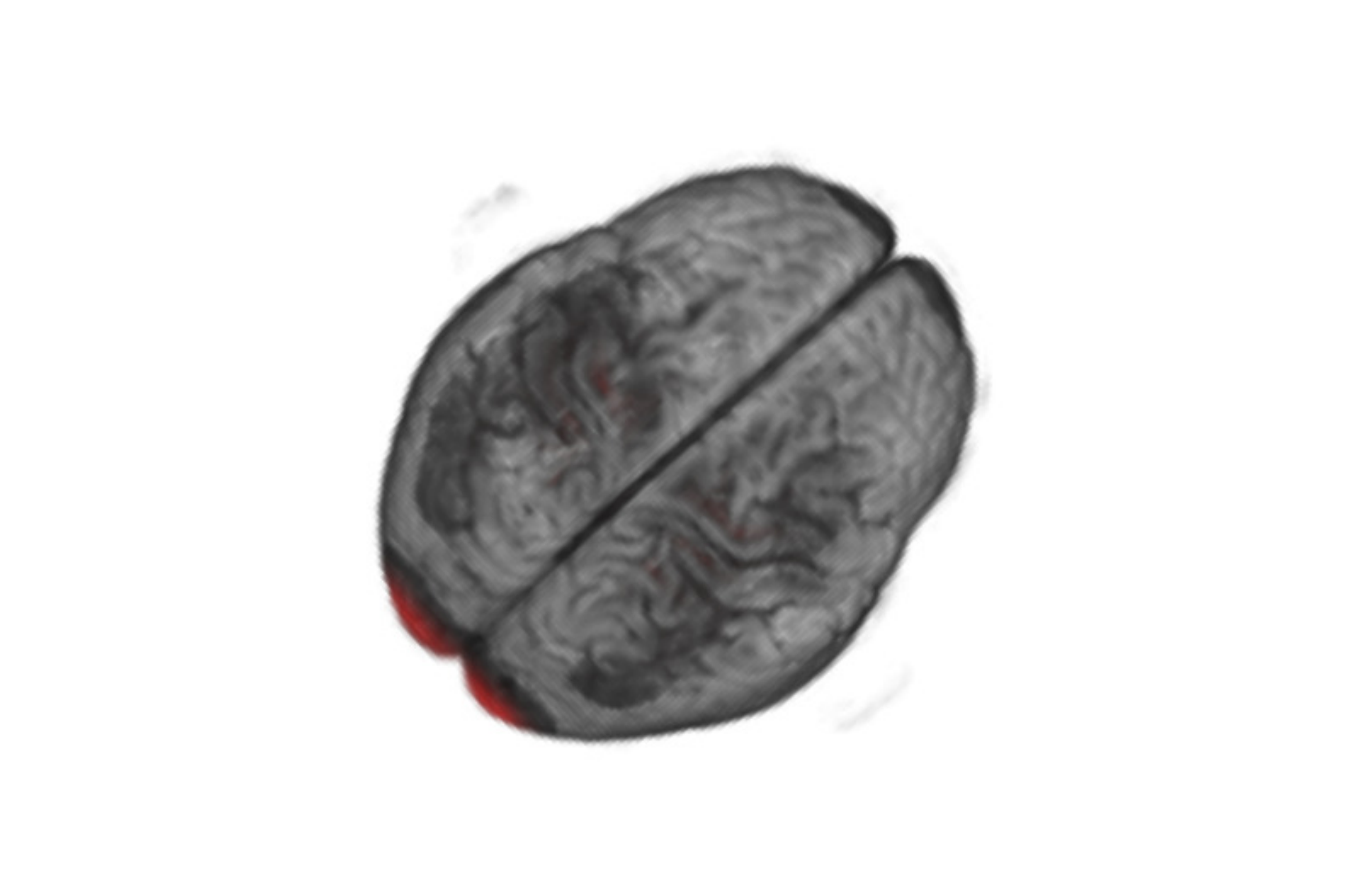} \\
\includegraphics[bb=220bp 0bp 750bp 520bp,scale=0.2]{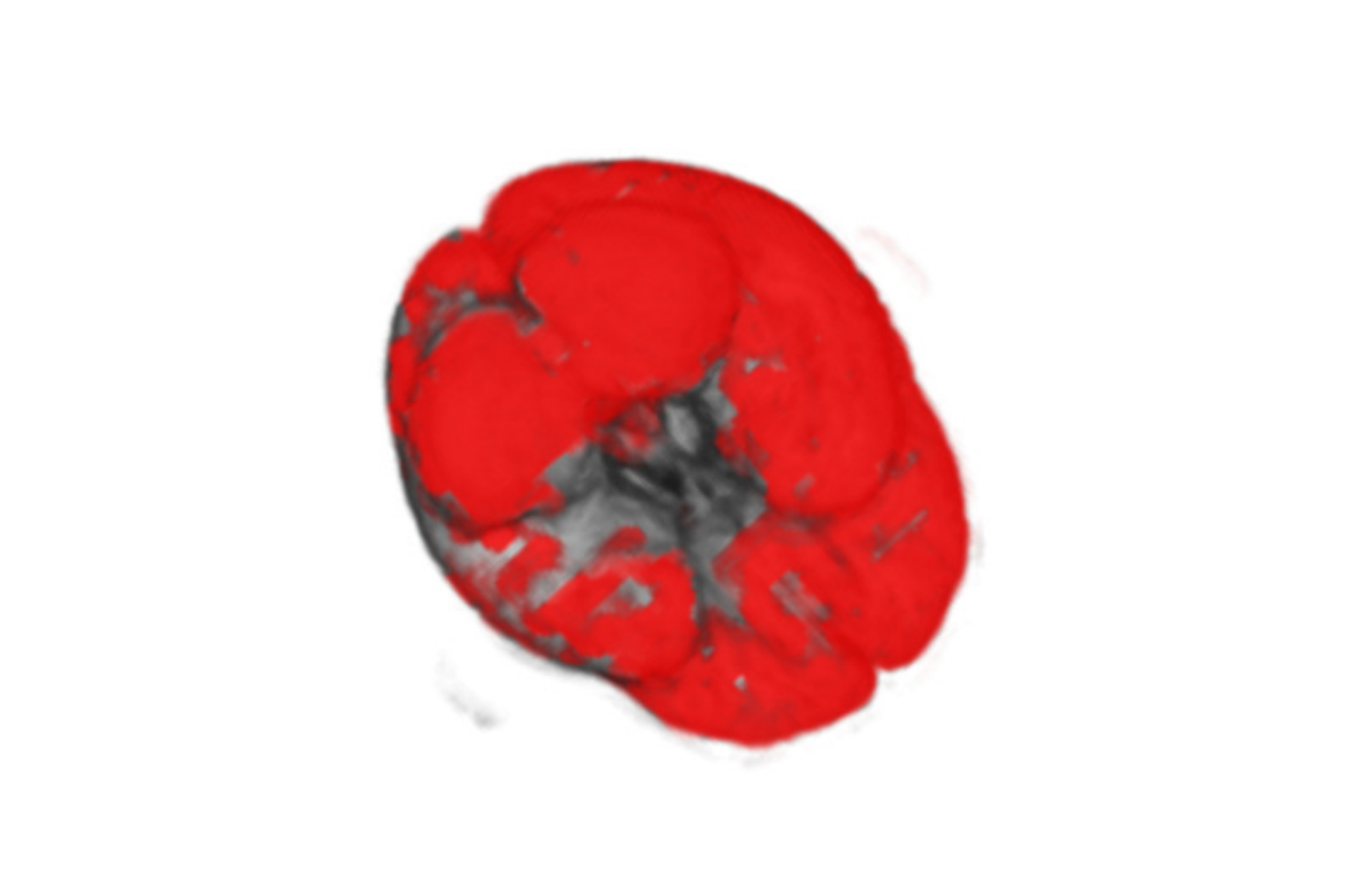}     &
\includegraphics[bb=220bp 0bp 750bp 520bp,scale=0.2]{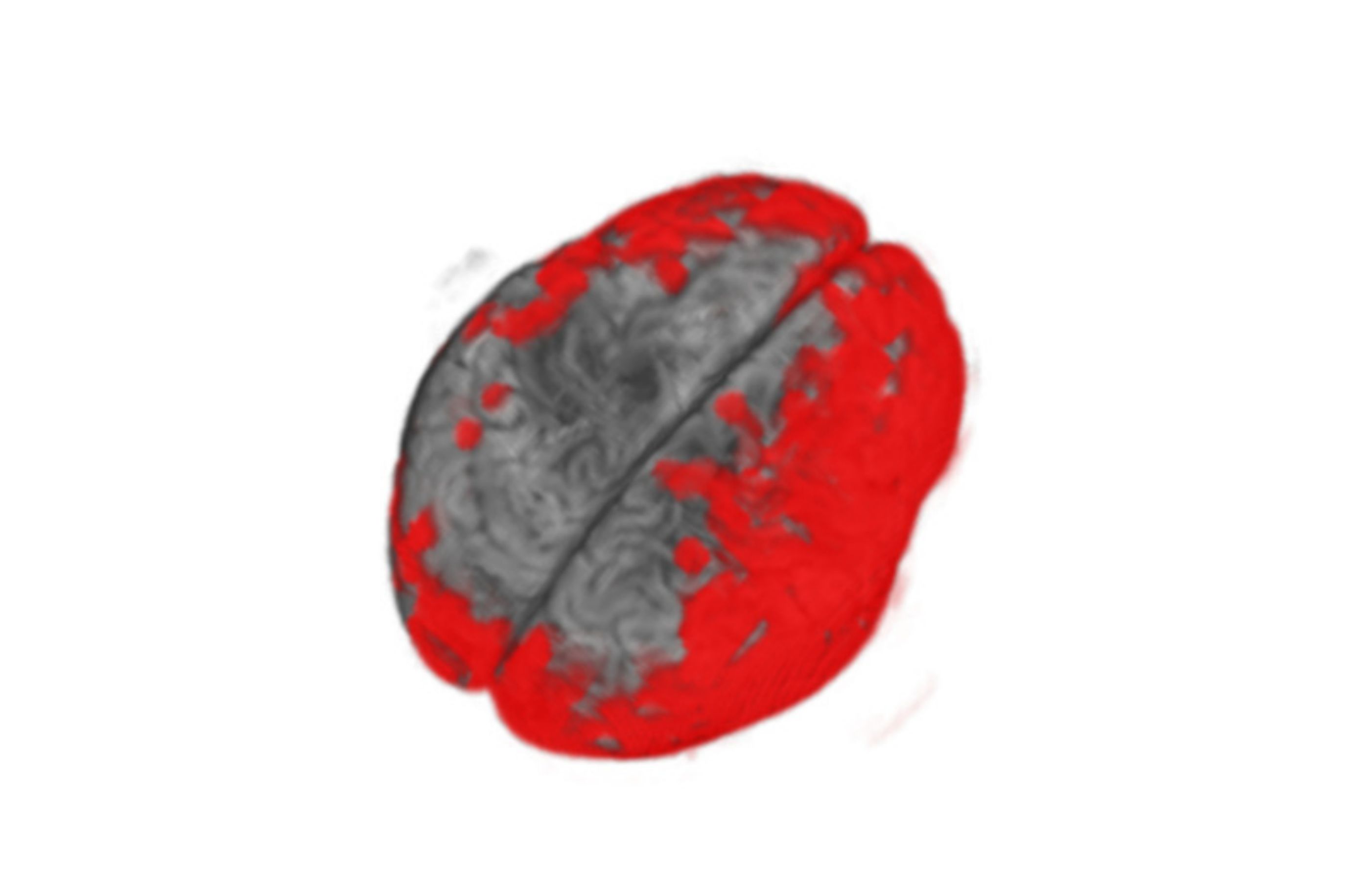}   & \mbox{ } &
\includegraphics[bb=220bp 0bp 750bp 520bp,scale=0.2]{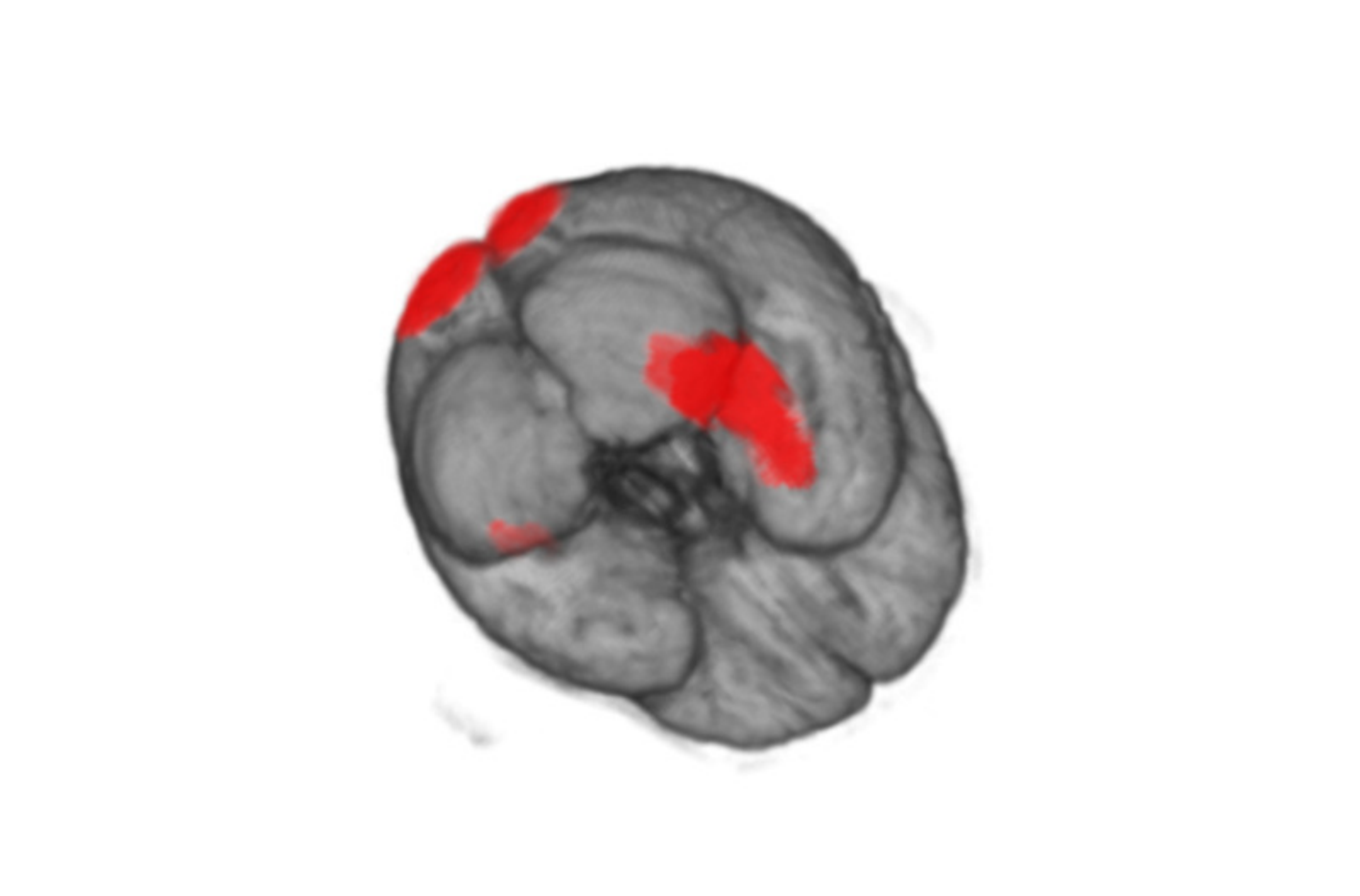}    &
\includegraphics[bb=220bp 0bp 750bp 520bp,scale=0.2]{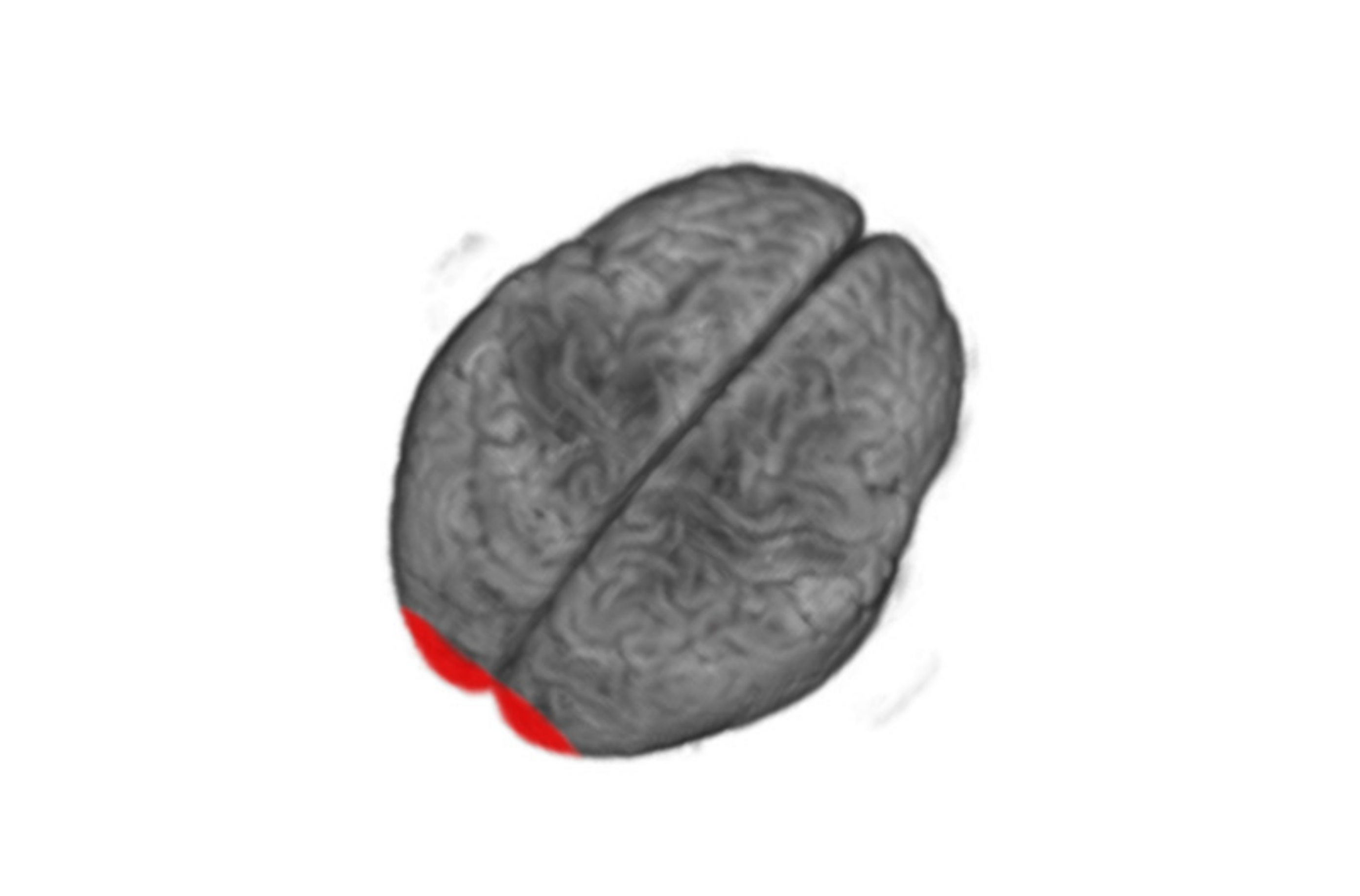}  \\
\end{tabular}
\end{center}
\vspace{-0.2in}
\caption{ADHD analysis: top panels are estimated regression coefficients; bottom panels are the $p$-value maps, where the red regions correspond to the $p$-value less than the threshold 0.05. The left four plots show the OLS estimation, and the right four the envelope estimation. For each estimator, two angles of views are shown. \label{fig:adhd}}
\end{figure}

We next analyzed a magnetic resonance imaging (MRI) data for a study of attention deficit hyperactivity disorder (ADHD). The data was produced by the ADHD-200 Sample Initiative, then preprocessed by the Neuro Bereau and made available at \url{http://neurobureau.projects.nitrc.org/ADHD200/Data.html}. It consists of 776 subjects, among whom 285 are combined ADHD subjects and 491 are normal controls. We removed 47 subjects due to the missing observations or poor image quality, then downsized the MRI images from $256\times198\times256$ to $30\times36\times30$, which is to serve as our 3-way tensor response. The predictors include the group indicator (1 for ADHD and 0 for control), the subject's age and gender. Figure~\ref{fig:adhd} summarizes the findings. While the OLS estimator reveals essentially no useful information, the envelope estimator shows clearly two regions that reflect distinctive activity pattern between the ADHD and control subjects. One region corresponds to the cuneus and the other to the fusiform gyrus, and such findings are consistent with the literature \citep{Booth2005,Solanto2009,Wolf2009,Zang2007,Tian2008}.

\section{Discussion}
\label{sec:discussion}

In this article, we have proposed a parsimonious model for regression with a tensor response and a vector of predictors. Adopting a generalized sparsity principle, we have developed an envelope based estimator that can identify and focus on the material information of the tensor response. By doing so, the number of free parameters is effectively reduced, and the resulting estimator is asymptotically efficient. Both simulations and real data analysis have demonstrated effectiveness of the new estimator.

We make some remarks about practical use of our proposed method. First, we suggest to combine the coefficient map and the $p$-value map in practice to help identify relevant signal regions. We have observed that the $p$-value map using the OLS asymptotic covariance can be conservative especially when the true signal is weak, whereas the coefficient map can often provide a useful recovery. On the other hand, the coefficient map may include many small signal regions, while the $p$-value map is usually more clean. Second, the working envelope dimension $u_k$ is the main tuning parameter in our proposal. In principle, if the selected working dimension is smaller than the truth, the corresponding envelope estimator is biased, whereas if the selected dimension is larger than the truth, the resulting estimator is unbiased but can be more variable. The selection of envelope dimension reflects a bias and variance trade-off.

A core idea of our proposal is to recognize and focus the estimation based upon the relevant information in the tensor response. Sparsity is defined in a general sense and is achieved through the envelope method. This is different from the common strategy in sparse modeling that induces sparsity through penalty functions. On the other hand, our envelope based estimation can be naturally \emph{coupled} with penalty functions to attain further regularization. This line of work is currently under investigation and consists of our future research.

\baselineskip=12pt
\bibliography{ref-tensor-env}
\bibliographystyle{apalike}

\end{document}